\documentclass[aps,twocolumn,twoside,superscriptaddress,longbibliography,nofootinbib, floatfix]{revtex4-1}

\usepackage{times}
\usepackage[utf8]{inputenc}
\usepackage{tikz}
\usetikzlibrary{positioning, decorations.pathmorphing, decorations.markings}
\usepackage{pgfplots}
\pgfplotsset{compat=newest}
\usepgfplotslibrary{fillbetween}
\usepackage[pdftex,colorlinks=true,linkcolor=blue,citecolor=blue,urlcolor=blue]{hyperref}
\usepackage{amsmath}
\usepackage{amssymb}
\usepackage{qcircuit}
\usepackage{comment}
\usepackage{bm}

\newcommand{\ket}[1]{|#1\rangle}
\newcommand{\bra}[1]{\langle#1|}
\newcommand{\braket}[1]{\langle#1\rangle}
\newcommand{\proj}[1]{|#1\rangle\langle#1|}

\newcommand{\be}{\begin{equation}}
\newcommand{\ee}{\end{equation}}

\newcommand{\C}{\mathbb{C}}
\newcommand{\Z}{\mathbb{Z}}

\newcommand{\tr}{\operatorname{tr}}

\begin{document}

\title{Strategies for solving the Fermi-Hubbard model on near-term quantum computers}
\author{Chris Cade}
\thanks{Present address: QuSoft and CWI, Amsterdam.}
\affiliation{Phasecraft Ltd.}
\author{Lana Mineh}
\affiliation{Phasecraft Ltd.}
\affiliation{School of Mathematics, University of Bristol}
\affiliation{Quantum Engineering Centre for Doctoral Training, University of Bristol}
\author{Ashley Montanaro}
\affiliation{Phasecraft Ltd.}
\affiliation{School of Mathematics, University of Bristol}
\author{Stasja Stanisic}
\affiliation{Phasecraft Ltd.}

\begin{abstract}
The Fermi-Hubbard model is of fundamental importance in condensed-matter physics, yet is extremely challenging to solve numerically. Finding the ground state of the Hubbard model using variational methods has been predicted to be one of the first applications of near-term quantum computers. Here we carry out a detailed analysis and optimisation of the complexity of variational quantum algorithms for finding the ground state of the Hubbard model, including costs associated with mapping to a real-world hardware platform. The depth complexities we find are substantially lower than previous work. We performed extensive numerical experiments for systems with up to 12 sites. The results suggest that the variational ans\"atze we used -- an efficient variant of the Hamiltonian Variational ansatz and a novel generalisation thereof -- will be able to find the ground state of the Hubbard model with high fidelity in relatively low quantum circuit depth. Our experiments include the effect of realistic measurements and depolarising noise. If our numerical results on small lattice sizes are representative of the somewhat larger lattices accessible to near-term quantum hardware, they suggest that optimising over quantum circuits with a gate depth less than a thousand could be sufficient to solve instances of the Hubbard model beyond the capacity of classical exact diagonalisation.
\end{abstract}

\date{\today}

\maketitle

\setlength{\parskip}{3pt}

Modelling quantum-mechanical systems is widely expected to be one of the most important applications of near-term quantum computing hardware~\cite{cirac12,georgescu14,preskill18}. Quantum computers could enable the solution of problems in the domains of many-body quantum physics and quantum chemistry that are intractable for today's best supercomputers.

Quantum algorithms have been proposed for both dynamic and static simulation of quantum systems. In the former case, one seeks to approximate time-evolution according to a certain quantum Hamiltonian. In many physically relevant cases, such as Hamiltonians obeying a locality constraint on their interactions, this can be carried out efficiently, i.e.\ in time polynomial in the system size~\cite{lloyd96}; by contrast, even to write down a classical description of the quantum system would take exponential time. However, in cases where the performance of the quantum simulation algorithm has been calculated and optimised in detail, solving a large enough problem instance to be practically relevant is still beyond the capabilities of present-day quantum computing technology. For example, several recent works describing highly-optimised algorithms for time-dynamics simulation~\cite{childs18,babbush18,nam19} determine complexities in the range of $10^5-10^8$ quantum gates to simulate systems beyond classical capabilities. By comparison, the most complex quantum circuit executed in the recent demonstration by Google of a quantum computation outperforming a classical supercomputer contained 430 two-qubit gates~\cite{sycamore}.

In the case of static simulation, the canonical problem is to produce the ground state of a quantum Hamiltonian. Once this state is produced, measurements can be performed to determine its properties. Although this problem is expected to be computationally hard for quantum computers in the worst case~\cite{gharibian15}, it is plausible that instances of practical importance could nevertheless be solved efficiently. A prominent class of methods for producing ground states are variational methods, and in particular the variational quantum eigensolver~\cite{peruzzo14,mcclean16} (VQE).
The VQE framework can be seen as a hybrid quantum-classical approach to produce a ground state of a quantum Hamiltonian $H$. A classical optimiser is used to optimise over quantum circuits which produce states $\ket{\psi}$ that are intended to be the ground state of $H$. The cost function provided to the optimiser is an approximation of the energy $\braket{\psi|H|\psi}$, which is estimated using a quantum computer.

Here our focus is on variational algorithms for a specific task: constructing the ground state of the iconic 2D Fermi--Hubbard model~\cite{hubbard63,hubbard13}. This model is of particular interest for several reasons. First, despite its apparent simplicity, its theoretical properties are far from fully understood~\cite{scalapino07,hubbard13,leblanc15}. Second, it is believed to be relevant to physical phenomena of extreme practical importance, such as high-temperature superconductivity~\cite{dagotto94}. Third, its regular structure and relatively simple form suggest that it may be easier to implement on a near-term quantum computer than, for example, model systems occurring in quantum chemistry.

The Hubbard Hamiltonian is defined as
\be \label{eq:hubbard} H = -t \sum_{\langle i, j \rangle,\sigma} (a_{i\sigma}^\dag a_{j\sigma} + a_{j\sigma}^\dag a_{i\sigma}) + U \sum_i n_{i\uparrow}n_{i\downarrow}, \ee
where $a_{i\sigma}^\dag$, $a_{i\sigma}$ are fermionic creation and annihilation operators; $n_{i\uparrow} = a_{i\uparrow}^\dag a_{i\uparrow}$ and similarly for $n_{i\downarrow}$; the notation $\langle i, j \rangle$ in the first sum associates sites that are adjacent in an $n_x \times n_y$ rectangular lattice (``grid''); and $\sigma \in \{\uparrow,\downarrow\}$. The first term in (\ref{eq:hubbard}) is called the hopping term with $t$ being the tunnelling amplitude, and the second term is called the interaction or onsite term where $U$ is the Coulomb potential. We will usually fix $t=1$, $U=2$ (similarly to~\cite{wecker15}); see Appendix \ref{sec:U_param} for results suggesting that the complexity of approximately finding the ground state of $H$ is not substantially different for other $U$ not too large and sufficiently bounded away from 0. We sometimes also consider what we call the non-interacting version of the Hubbard model, which only contains the hopping term.

On an $n_x \times n_y$ grid, the Hubbard Hamiltonian can be represented as a sparse square matrix with $2^{2n_x n_y}$ rows. Although the size of this matrix can be reduced by restricting to a subspace corresponding to a given occupation number, and taking advantage of translation- and spin-invariance, the worst-case growth of the size of these subspaces is still exponential in $N = n_xn_y$. This exponential growth severely limits the capability of classical exact solvers to address this model. For example, Yamada, Imamura and Machida~\cite{yamada05} report an exact solution of the Hubbard model with 17 fermions on 22 sites requiring over 7TB of memory and 13 TFlops on a 512-node supercomputer. By contrast, a Hubbard model instance with $N$ sites can be represented using a quantum computer with $2N$ qubits (each site can contain at most one spin-up and at most one spin-down fermion, so 2 qubits are required per site). This suggests that a quantum computer with around 50 qubits could already simulate instances of the Hubbard model going beyond classical capabilities.

Approximate classical techniques such as the quantum Monte Carlo and Density Matrix Renormalisation Group methods can address larger grids (up to thousands of sites) than near-term quantum computers, but experience difficulties in certain coupling regimes and away from half-filling, leading to substantial uncertainties in physical quantities~\cite{leblanc15}. The hope is that quantum computing, while addressing smaller system sizes, could evade the difficulties experienced by these methods (such as the ``sign problem'' in quantum Monte Carlo methods) and enable access to these regimes.

Another approach to understanding the Hubbard model via a quantum device is analogue quantum simulation~\cite{georgescu14,altman19}: engineering a special-purpose quantum system that implements the Hubbard Hamiltonian directly~\cite{hensgens17,tarruell18,gross17}. Analogue quantum simulators are easier to implement experimentally than universal quantum computers, and enable access to much larger systems than will be possible using near-term quantum computers. However, they are inherently less flexible than digital quantum simulation in terms of the Hamiltonians that can be implemented and the measurements that can be performed, and experience difficulties with reaching sufficiently low temperatures to demonstrate phenomena such as superconductivity~\cite{esslinger10,tarruell18,altman19}.

Prior work on variational methods for solving the Hubbard model~\cite{wecker15,reiner19,Verdon2019,Dallaire-Demers2018} (discussed in Section \ref{sec:variational_method}) has left a number of important questions open which must be answered to understand whether it is a realistic target for near-term quantum computers. These include: what is the precise complexity of implementing the variational ansatz? How well will the optimisation routines used handle statistical noise, and noise in the quantum circuit? How complex is the procedure required to produce the initial state? 

Here we address all these questions and develop detailed resource estimates and circuit optimisations, as well as extensive numerical experiments for grids with up to 12 sites (24 qubits), in order to estimate how well realistic near-term quantum computers will be able to solve the Hubbard model. Although the Hubbard model is easily solvable directly by a classical algorithm for systems of this size, these experiments give insight into the likely performance of VQE on instances that are beyond this regime. Unlike some previous work, our focus is on solving instances just beyond the capability of classical hardware (e.g.\ size $10 \times 10$ or smaller) using machines with few (e.g.\ at most 200) physical qubits. In this regime, it is essential to carry out precise complexity calculations to understand the feasibility of the VQE approach.

A key ingredient in the complexity calculations for our circuits will be their depths. To compute this, we assume that the quantum computer can implement arbitrary 2-qubit gates, and that 1-qubit gates can be implemented at zero cost. These assumptions are not too unrealistic. Almost all the 2-qubit gates we will need are rotations of the form $e^{i(\theta(XX+YY) + \gamma ZZ)}$ (up to single-qubit unitaries), which can be implemented natively on some superconducting qubit platforms; and 1-qubit gates can be implemented at substantially lower cost in some architectures~\cite{barends14}.

When simulating a VQE experiment on a classical computer, one can consider three different levels of realism:
\begin{itemize}
    \item The simplest but least realistic level is to assume that we can perform exact energy measurements to learn $\braket{\psi|H|\psi}$, which can be used directly as input to a classical optimiser.
    
    \item The next level of realism is to simulate the result of energy measurements as if they were performed on a quantum computer, but to assume that the quantum computer is perfect, i.e.\ does not experience any noise.
    
    \item Finally, one can simulate the effect of noise during the quantum computation.
\end{itemize}
In this work we consider all of these levels. The main results we obtain can be summarised as follows: 
\begin{itemize}
\item The most efficient approach we found for encoding fermions as qubits, for the small-sized grids we consider (indeed, for grids such that $\min\{n_x,n_y\} \le 8$), was the Jordan-Wigner transform, both in terms of space and (perhaps surprisingly) in terms of circuit depth. See Appendix \ref{app:otherencodings} for details.
\item We develop an approach to efficiently implement a variant of the so-called ``Hamiltonian variational'' (HV) ansatz~\cite{wecker15}, and generalisations of this ansatz, in the Jordan-Wigner transform (Section \ref{sec:efficient_np}). The circuit depth is as low as $2n_x + 1$ per ansatz layer on a fully-connected architecture, and $6n_x + 1$ per layer on an architecture such as Google Sycamore~\cite{sycamore}. See Table \ref{tab:circuit_depths} for some examples. This method can also be used to implement the fermionic Fourier transform (FFT) more efficiently than previous work for small grid sizes.

\item We introduce an efficient method of measuring the energy of a trial state produced using this ansatz (Section \ref{sec:measurement}), which requires only 5 computational basis measurements and allows for a simple notion of error-detection.
\item In numerical experiments with simulated exact energy measurements and using the L-BFGS optimiser, the error with the true ground state (measured either by fidelity or energy error) decreases exponentially with the circuit depth in layers (Figure \ref{fig:depthscaling}). This gives good evidence that the efficient HV ansatz is able to represent the ground state of the Hubbard model efficiently, at least for the small grid sizes accessible to near-term hardware.
\item For all grids with at most 12 sites, 0.99 fidelity to the ground state (which is non-degenerate in all the cases we consider) can be achieved using an efficient HV ansatz circuit with at most 18 layers (Figure \ref{fig:repdepths}). The results are consistent with a grid with $N$ sites needing $O(N)$ layers; in all cases, we found that at most $1.5N$ layers were needed. We present a generalisation of the HV ansatz called the Number Preserving ansatz (Section \ref{sec:var_ansatz}), which gives more freedom in the choice of gates. This generally performs better in terms of the depth required to achieve high fidelity with the ground state, but requires more optimisation steps.
\item In numerical experiments with simulated realistic energy measurements on systems with up to nine sites, the coordinate descent~\cite{nakanishi19,ostaszewski19,parrish19} and SPSA~\cite{spsa,kandala17, Ganzhorn2019} algorithms are both able to achieve high fidelity with the ground state (e.g.\ SPSA achieves fidelity $>0.977$ for a $3\times 3$ grid; see Table \ref{tab:realisticfidelities} and Figure \ref{fig:realistic}) by making a number of measurements which would require a few hours\footnote{$\sim 57$M circuit evaluations; Google's Sycamore processor can perform 1M circuit evaluations in 200s~\cite{sycamore}.} of execution time on a real quantum computer. On $2\times 2$ and $2\times 3$ grids the two algorithms achieved similar final fidelities, while on $1\times 6$ and $3\times 3$ grids SPSA performed substantially better.
\item In numerical experiments with simulated depolarising noise in the quantum circuit for systems with up to 6 sites, error rates of up to $10^{-3}$ do not have a significant effect on the fidelity of the solution (Table~\ref{tab:noisyresults}). The use of error-detection gives a small but noticeable improvement to the fidelity (Figure~\ref{fig:noisy2x3}).
\end{itemize}
\begin{table}[t]
\centering
\begin{tabular}{|c|c|}
\hline 
Architecture & Ansatz circuit depth per layer
\\ \hline
Fully Connected       & \begin{tabular}[c]{@{}c@{}} {$\bf 2n_x+1$ / $\bf 2n_x+2$}\\ $4\times 4 \,/\, 4\times 5 : 9$\\ $5\times 5 \,/\, 5\times 6: 12$\\ $6 \times 6 : 13$\end{tabular}   \\
\hline
Nearest Neighbour     & \begin{tabular}[c]{@{}c@{}}{$\bf 4n_x$} / {$\bf 4n_x+1$}\\ $4\times 4 \,/\, 4\times 5: 16 $\\ $5\times 5 \,/\, 5\times 6: 21$\\ $6 \times 6 : 24$\end{tabular} \\
\hline
Google Sycamore     & \begin{tabular}[c]{@{}c@{}}{$\bf 6n_x+1$} / {$\bf 6n_x+2$}\\ $4\times 4 \,/\, 4\times 5 : 25$\\ $5\times 5 \,/\, 5\times 6: 32$\\ $6 \times 6 : 37$\end{tabular}
\\ \hline
\end{tabular}
\caption{Example circuit depths per layer of the efficient ans\"atze for various architectures (for $n_x$ even/odd).}
\label{tab:circuit_depths}
\end{table}
We conclude that variational methods show significant promise for producing the ground state of the Hubbard model for grid sizes somewhat beyond what is accessible with classical computational methods. Highly-optimised ansatz circuits can be designed; the depth required for these circuits to find the ground state seems to scale favourably with the size of grid; and the use of realistic measurements and noise in the circuit do not reduce final fidelities unreasonably.

Based on these results, it seems plausible that an instance of the Hubbard model larger than the capacity of classical exact diagonalisation methods could be solved by optimising over quantum circuits with depth 300--500 (on a fully-connected architecture). This is substantially smaller than previous estimates for other proposed applications of near-term quantum computers, albeit beyond the capacity of leading hardware available today. Although exact diagonalisation provides more information than producing the ground state on a quantum computer, physically important quantities (such as correlation functions) are nevertheless accessible. This suggests that variational quantum algorithms could become an important tool for the study of the Hubbard model.


\section{The variational method}
\label{sec:variational_method}

Our work fits within the standard VQE framework~\cite{peruzzo14,mcclean16}. The field of variational quantum algorithms is already too vast to sensibly summarise here. The VQE algorithm has been implemented experimentally in a number of platforms including photonics~\cite{peruzzo14},  superconducting qubits~\cite{kandala17, OMalley2016, Ganzhorn2019} and trapped ions~\cite{Hempel2018, Shen2017, Nam2019, Shehab2019}, while there have also been numerous theoretical developments~\cite{Moll2018,Wecker2014, wecker15, Dallaire-Demers2018, Grimsley2019, Gard2019,Lee2018, Barkoutsos2018, Romero17}.

A number of works have applied VQE to the Hubbard model specifically. Wecker et al.~\cite{wecker15} developed the Hamiltonian Variational (HV) ansatz, which will be a key tool that we will use and expand upon (see Section \ref{sec:var_ansatz}).
They tested it for the half-filled Hubbard model for systems of up to 12 sites -- in the case of simulated exact energy measurements, they used ladders with dimensions $n_x \times 2$ for $n_x = 2,\dots,6$; in the case of realistic energy measurements, they tested a system of size $4 \times 2$. Implementation of 2 layers of this ansatz for a $4 \times 2$ system would require 1000 gates according to their estimate (we reduce this estimate substantially; see Section \ref{sec:implementation}). Dallaire-Demers et al.~\cite{Dallaire-Demers2018} have also developed a low-depth circuit ansatz inspired by the unitary coupled cluster ansatz and applied it to the $2\times 2$ Hubbard model.

Reiner et al.~\cite{reiner19} have recently studied how gate errors affect the HV ansatz. They considered a model where gates are subject to fixed unitary over-rotation errors, and found that for small system sizes (grids of size $2\times 2$, $3\times 2$ and $3 \times 3$), reasonably small errors did not prevent the variational algorithm from finding a high-quality solution.
Verdon et al.~\cite{Verdon2019} developed an approach to optimising VQE parameters using recurrent neural networks, and applied it to Hubbard model instances of size $2\times 2$, $3 \times 2$ and $4\times 2$. Wilson et al.~\cite{wilson19} designed a somewhat related ``meta-learning'' approach to VQE which they tested on the {\em spinless} Hubbard model on 3 sites.

We also remark that several endeavours (e.g.~\cite{dallaire-demers16,dallaire-demers16a,reiner18,babbush18}) have studied the complexity of quantum algorithms for simulating time-evolution or thermodynamic properties of the Hubbard model.

The VQE framework requires a few different ingredients to be specified:
\begin{enumerate}
    \item The encoding used to represent fermions as qubits
    \item The properties of variational ansatz (circuit family, initial state, etc.)
    \item Implementation of energy measurements
    \item Selection of classical optimiser
\end{enumerate}
Additionally, there are some important implementation details to be determined for the resulting quantum circuits to be executed in a real-world architecture. In the remainder of this section, we describe the approach we took to fill in all these details.


\subsection{Fermionic encoding}\label{sec:ferm_encoding}

We use the well-known Jordan-Wigner encoding of the fermionic Hamiltonian $H$ as a qubit Hamiltonian. This encoding has no overhead in qubit count, as each site maps to two qubits. The downside is that some fermionic interactions map to long strings of Pauli operators, whose length increases with the grid size. We will need to implement time-evolution according to the hopping terms in $H$; this also has complexity that increases with the grid size.

There are other encodings (such as the Bravyi-Kitaev super-fast encoding~\cite{bravyi02} and Ball-Verstraete-Cirac encoding~\cite{ball05,verstraete05}) which produce local operators, at the expense of using additional qubits. However, for small grid sizes, the complexity of the corresponding quantum circuits for time-evolution seems to be higher than optimised methods that use fermionic swap networks to implement the required time-evolution operations under the Jordan-Wigner transform. See Appendix \ref{app:otherencodings} for a discussion.

\begin{figure}[t]
    \centering
    \includegraphics[scale=0.6]{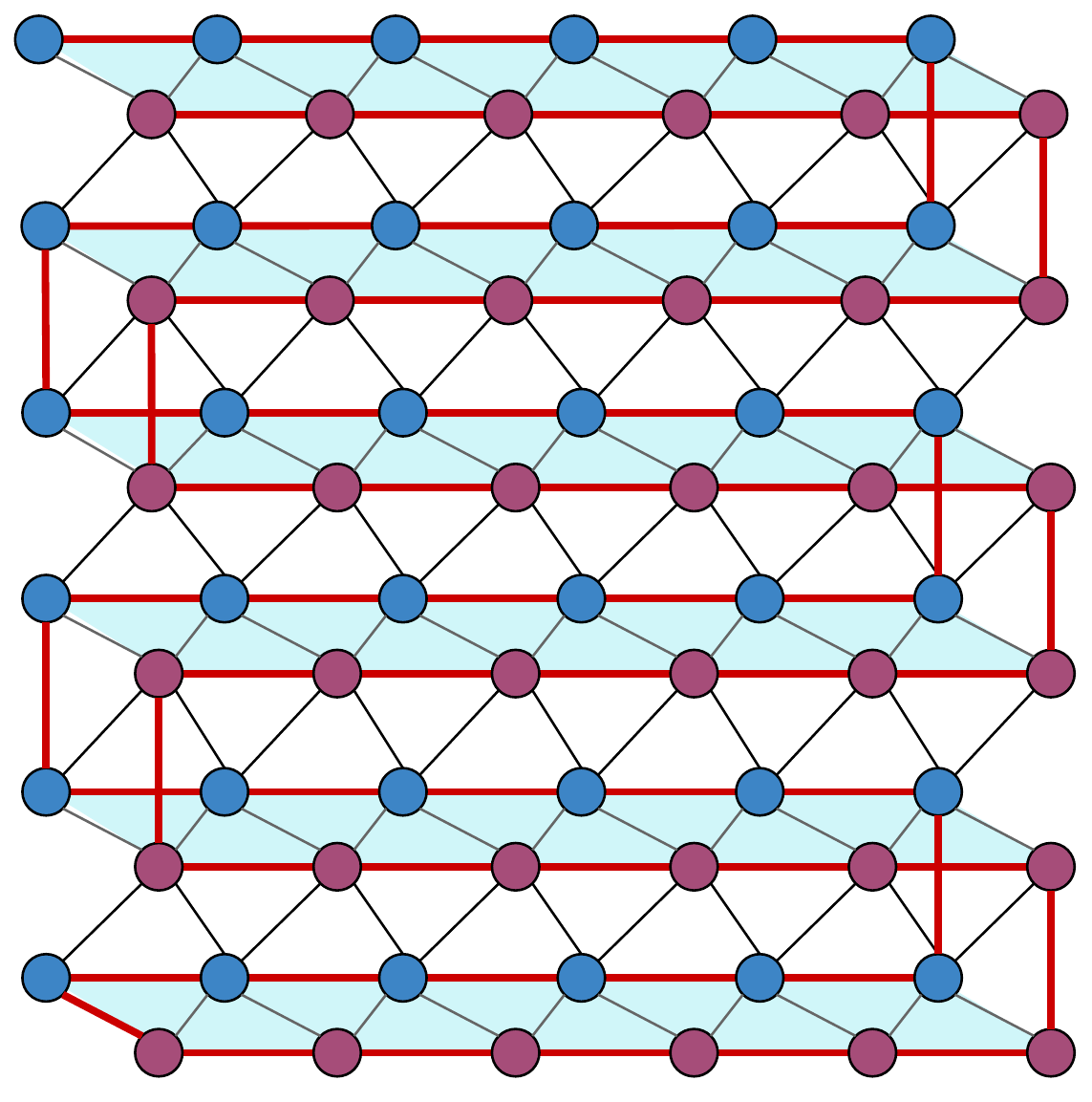}
    \caption{An illustration of how fermionic modes can be mapped to physical qubits on a physical architecture such as Google's Sycamore device~\cite{sycamore}. The fermionic modes (blue: spin-up, red: spin-down) on a $6\times 6$ lattice are mapped to qubits in an array of size $2\times6\times6$. The red line represents the order associated with the JW encoding of the qubits, which moves from the top left towards the right. The blue panels are added to aid visualisation. Note that the red line does not follow the true connectivity of the qubits (the thin black lines), and hence any `local' operator with respect to the JW encoding is not necessarily local with respect to the physical connectivity of the qubits, and vice versa.}
    \label{fig:snake_shape}
\end{figure}

The Jordan-Wigner encoding associates each fermionic mode (corresponding to a site on a grid and a choice of spin) with a qubit. The encoding can be seen as assigning a position on a line to each fermionic mode. We use the so-called `snake-shaped' configuration shown in Figure \ref{fig:snake_shape}, which illustrates a setting where the qubits are laid out according to the Google Sycamore architecture~\cite{sycamore}\footnote{That is, a natural generalisation of the qubit topology reported in~\cite{sycamore} to larger system sizes.}. The advantage of using this configuration is that we can make use of fermionic swap networks for efficiently implementing the ansatz circuits (see section \ref{sec:efficient_np}) and carry out Hamiltonian measurements using the lowest number of circuit preparations (see section \ref{sec:measurement}). 

Each hopping term between qubits $i$ and $j$ ($i < j$) maps to a qubit operator via
\[ a_i^\dag a_j + a_j^\dag a_i \mapsto \frac{1}{2}(X_iX_j + Y_iY_j) Z_{i+1} \cdots Z_{j-1}. \]
For $j=i+1$ (a hopping term between horizontally adjacent qubits), there is only the `bare hopping term' $\frac{1}{2}(X_iX_j + Y_iY_j)$. For vertically adjacent qubits, the bare hopping term is accompanied by the string of $Z$ operators $Z_{i+1} \cdots Z_{j-1}$.   Each onsite term acting on qubits $i$ and $j$ maps to a qubit operator via 
\[ a_i^\dag a_i a_j^\dag a_j \mapsto \frac{1}{4}(I - Z_i)(I - Z_j), \]
whether or not qubits $i$ and $j$ are adjacent in the Jordan-Wigner encoding.
Hence, as we will see, the vertical hopping terms are the most difficult of these three types of terms to implement efficiently.

\subsection{Variational ans\"atze}\label{sec:var_ansatz}

Various variational ans\"atze have been proposed for use within the VQE framework, including the Hamiltonian variational (HV) ansatz~\cite{wecker15}, hardware-efficient ans\"atze \cite{kandala17}, unitary coupled cluster~\cite{peruzzo14, Shen2017}, and others.

The HV ansatz is based on intuition from the quantum adiabatic theorem, which states that one can evolve from the ground state of a Hamiltonian $H_A$ to the ground state of another Hamiltonian $H_B$ by applying a sequence of evolutions of the form $e^{-it H_A}$, $e^{-it H_B}$ for sufficiently small $t$. In the case of the Hubbard model, we start in the ground state of the non-interacting Hubbard Hamiltonian ($U=0$) for a given occupation number, which can be prepared efficiently~\cite{verstraete2009quantum,jiang2018quantum}, and then evolve to the ground state of the full Hubbard model, including the onsite terms.

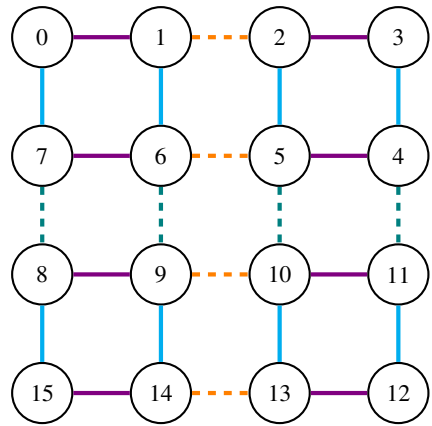
\begin{figure}[t]
    \centering
    \begin{tikzpicture}[
    roundnode/.style={circle, draw=black, thick, minimum size=8mm, scale=1}, node distance = 0.75cm
    ]
    \node[roundnode]    (qubit0) {0};
    \node[roundnode]    (qubit1)    [right = of qubit0] {1};
    \node[roundnode]    (qubit2)    [right = of qubit1] {2};
    \node[roundnode]    (qubit3)    [right = of qubit2] {3};
    \node[roundnode]    (qubit7)    [below = of qubit0] {7};
    \node[roundnode]    (qubit6)    [right = of qubit7] {6};
    \node[roundnode]    (qubit5)    [right = of qubit6] {5};
    \node[roundnode]    (qubit4)    [right = of qubit5] {4};
    \node[roundnode]    (qubit8)    [below = of qubit7] {8};
    \node[roundnode]    (qubit9)    [right = of qubit8] {9};
    \node[roundnode]    (qubit10)    [right = of qubit9] {10};
    \node[roundnode]    (qubit11)    [right = of qubit10] {11};
    \node[roundnode]    (qubit15)    [below = of qubit8] {15};
    \node[roundnode]    (qubit14)    [right = of qubit15] {14};
    \node[roundnode]    (qubit13)    [right = of qubit14] {13};
    \node[roundnode]    (qubit12)    [right = of qubit13] {12};
    
    \draw[-, ultra thick, draw=violet] (qubit0.east) -- (qubit1.west);
    \draw[dashed, ultra thick, draw=orange] (qubit1.east) -- (qubit2.west);
    \draw[-, ultra thick, draw=violet] (qubit2.east) -- (qubit3.west);
    \draw[-, ultra thick, draw=violet] (qubit7.east) -- (qubit6.west);
    \draw[dashed, ultra thick, draw=orange] (qubit6.east) -- (qubit5.west);
    \draw[-, ultra thick, draw=violet] (qubit5.east) -- (qubit4.west);
    \draw[-, ultra thick, draw=violet] (qubit8.east) -- (qubit9.west);
    \draw[dashed, ultra thick, draw=orange] (qubit9.east) -- (qubit10.west);
    \draw[-, ultra thick, draw=violet] (qubit10.east) -- (qubit11.west);
    \draw[-, ultra thick, draw=violet] (qubit15.east) -- (qubit14.west);
    \draw[dashed, ultra thick, draw=orange] (qubit14.east) -- (qubit13.west);
    \draw[-, ultra thick, draw=violet] (qubit13.east) -- (qubit12.west);
    \draw[-, ultra thick, draw=cyan] (qubit0.south) -- (qubit7.north);
    \draw[-, ultra thick, draw=cyan] (qubit1.south) -- (qubit6.north);
    \draw[-, ultra thick, draw=cyan] (qubit2.south) -- (qubit5.north);
    \draw[-, ultra thick, draw=cyan] (qubit3.south) -- (qubit4.north);
    \draw[dashed, ultra thick, draw=teal] (qubit7.south) -- (qubit8.north);
    \draw[dashed, ultra thick, draw=teal] (qubit6.south) -- (qubit9.north);
    \draw[dashed, ultra thick, draw=teal] (qubit5.south) -- (qubit10.north);
    \draw[dashed, ultra thick, draw=teal] (qubit4.south) -- (qubit11.north);
    \draw[-, ultra thick, draw=cyan] (qubit8.south) -- (qubit15.north);
    \draw[-, ultra thick, draw=cyan] (qubit9.south) -- (qubit14.north);
    \draw[-, ultra thick, draw=cyan] (qubit10.south) -- (qubit13.north);
    \draw[-, ultra thick, draw=cyan] (qubit11.south) -- (qubit12.north);
    
    \end{tikzpicture}    
    
    \caption{The four sets of hopping terms (for a fixed spin). Hopping terms of the same colour commute, and hence in principle can be implemented simultaneously. Purple corresponds to the horizontal terms $H_1$, dashed orange to $H_2$, blue to the vertical terms $V_1$ and dashed green to $V_2$.}
    \label{fig:hopping_meas}
\end{figure}

Rather than alternating evolutions according to the full hopping and onsite terms in the Hamiltonian $H$ in (\ref{eq:hubbard}), it is natural to split $H$ into parts that consist of terms that are sums of commuting components, which could allow for more efficient time-evolution. This also allows for these terms to have different coefficients, while still respecting overall symmetries of the Hamiltonian. Then a layer of the HV ansatz is a unitary operator of the form
\be \label{eq:hv} e^{it_{V_2} H_{V_2}} e^{it_{H_2} H_{H_2}} e^{it_{V_1} H_{V_1}} e^{it_{H_1} H_{H_1}} e^{it_{H_O} H_O} \ee
where $H_O$ is the onsite term; $H_{V_1}$ and $H_{V_2}$ are the vertical hopping terms; $H_{H_1}$ and $H_{H_2}$ are the horizontal hopping terms as shown in Figure \ref{fig:hopping_meas}. Different layers can have different parameters. Note that there is some freedom in the order with which we can implement these terms, and also that some of them may not be needed depending on the grid dimensions. The vertical hopping terms are nontrivial to implement efficiently in the JW transform, given the potentially long strings of $Z$ operators associated with each of them. We remark that a similar technique of decomposition into commuting parts is common in quantum Monte Carlo methods, where it is known as the checkerboard decomposition.

The HV ansatz has been shown to be effective for small Hubbard model instances~\cite{wecker15,reiner19}, and involves a small number of variational parameters: at most 5 per layer.
One disadvantage of this ansatz is that preparing the initial state is a nontrivial task.
It can be produced using the (2D) fermionic Fourier transform (FFT), for which efficient algorithms are known~\cite{verstraete2009quantum,jiang2018quantum}, or via a direct method based on the use of Givens rotations~\cite{jiang2018quantum}. We calculated the complexity of an asymptotically fast algorithm for the FFT presented in~\cite{jiang2018quantum} and also developed an alternative implementation strategy using fermionic swap networks, which may be of independent interest. We found that, for grids of size up to $20 \times 20$, neither of these strategies was more efficient than direct preparation of the initial state using Givens rotations~\cite{jiang2018quantum}, which has circuit depth $n_x n_y - 1$ (assuming an arbitrary circuit topology). See Appendix \ref{app:fft} for the details.

To avoid this depth overhead for constructing the initial state, we also considered an ansatz which is a generalisation of HV. This ansatz benefits from the same theoretical guarantees that arbitrary-length circuits can find the ground state of $H$ while being more general and allowing for an initial state that is significantly more straightforward to generate. However, the trade-off is that it uses more parameters, making the optimisation process more challenging.

The ansatz, which we will call the Number Preserving (NP) ansatz, is derived from HV by replacing all hopping and onsite terms with a more general number-preserving operator\footnote{This is similar to the exchange-type entangling gates discussed in~\cite{Ganzhorn2019, Barkoutsos2018}; an alternative notion of number-preserving VQE ansatz was studied in~\cite{Gard2019}.} parameterised by two angles $\theta$ and $\phi$, and implemented by the 2-qubit unitary
\[U_{\text{NP}}(\theta, \phi) = 
    \begin{pmatrix}
    1 & 0 & 0 & 0 \\
    0 & \cos \theta & i \sin \theta & 0 \\
    0 & i \sin \theta & \cos \theta & 0 \\
    0 & 0 & 0 & e^{i\phi}
    \end{pmatrix}.
\]
The non-interacting ground state can still be used as the initial state, although computational basis states (where the Hamming weight is equal to the fermonic occupation number of interest) can also be used with some success (see Appendix \ref{app:init_state}). Then one layer of the ansatz consists of applying a $U_{\text{NP}}(\theta, \phi)$ gate (with varying angles $\theta$, $\phi$) across each pair of qubits that correspond to fermionic modes that interact according to the Hubbard Hamiltonian $H$ in (\ref{eq:hubbard}). That is, we apply $U_{\text{NP}}(\theta, \phi)$ gates for all pairs of modes $(i,\sigma)$, $(j,\tau)$ such that either $i \sim j$ and $\sigma = \tau$ (hopping terms), or $i = j$ and $\sigma \neq \tau$ (onsite terms). As before, different layers can have different parameters.

For an $n_x \times n_y$ grid, one layer of the NP ansatz requires
\[ 2 (2(n_x(n_y-1) + n_y(n_x-1)) + n_x n_y ) = 10n_x n_y - 4n_x - 4n_y\]
parameters. The HV ansatz is the special case of the NP ansatz that also preserves spin and where many parameters are fixed to be identical or 0.

\subsection{Efficient implementation of HV and NP ans\"atze}
\label{sec:efficient_np}
Hopping terms between vertically adjacent qubits that are not local with respect to the JW encoding must be accompanied by a string of $Z$ operators (see Section~\ref{sec:ferm_encoding}), which can be costly to implement. To reduce the overhead associated with these vertical hopping terms, we use a technique of Kivlichan et al.~\cite{kivlichan18} based on networks of fermionic SWAP gates, though with some minor changes for efficiency. In particular, we remove some unnecessary vertical fermionic swap gates and instead only swap horizontally adjacent qubits. This means that, for an $n \times n$ grid, only $n$ repetitions of a column-permuting subroutine (which itself has depth 2) are necessary to be able to implement all vertical hopping terms locally, in comparison to the $\frac{3}{\sqrt{2}}n$ iterations that are deemed to be necessary in~\cite{kivlichan18}. We now describe this approach; Appendix \ref{app:fft} gives a comparison to the approach of~\cite{kivlichan18}, in the closely analogous context of implementing the FFT. In what follows, we write `JW-adjacent' to mean `adjacent with respect to the JW encoding', and when we say that an operator is implemented locally, we mean that the two qubits that it acts on are JW-adjacent.

We use fermionic SWAP (FSWAP) gates to move qubits that were originally not JW-adjacent into JW-adjacent positions. The FSWAP gate acts as a SWAP gate for fermions, and corresponds to the unitary operator
\[ \begin{pmatrix} 1 & 0 & 0 & 0 \\ 0 & 0 & 1 & 0 \\ 0 & 1 & 0 & 0 \\ 0 & 0 & 0 & -1 \end{pmatrix}. \]
This allows vertical hopping interactions to be implemented locally, whilst maintaining the correct parity on all qubits. That is, we repeatedly apply the operator $U_RU_L$, where $U_L$ swaps odd-numbered columns with those to their right, and $U_R$ swaps even-numbered columns with those to their right. After each application of $U_RU_L$, a new set of qubits that were previously not vertically JW-adjacent are made JW-adjacent, meaning that the vertical hopping interaction between them can be implemented locally using a single number-preserving operator, without $Z$-strings. For an $n_x \times n_y$ grid, it suffices to apply $U_RU_L$ a total of $n_x$ times to allow all vertical interactions to be implemented locally and return the qubits to their original positions. 

\tikzset{qubit/.style={shape=circle,fill=black, scale=0.5}}
\tikzset{qubit2/.style={shape=circle, draw=violet, scale=1.2}}
\tikzset{->-/.style={decoration={markings, mark=at position #1 with {\arrow{latex}}},postaction={decorate}}}
\begin{figure}[t]
    \centering
    \begin{minipage}{\linewidth}
    \begin{tikzpicture}[node distance = 0.2cm]
        \foreach \x in {0,...,3}
        \foreach \y in {0,...,3}
        {
            \node[qubit] (\x-\y) at (\x,\y){};
        }
        
        \foreach \mynode/\joinednode in {3-0/3-1, 3-2/3-3, 0-1/0-2}
        {
            \draw[-, line width = 0.1cm, draw=cyan] (\mynode.north) -- (\joinednode.south) ;  
        }
        
        \foreach \mynode/\joinednode in {0-3/3-3, 0-2/3-2, 0-1/3-1, 0-0/3-0}
        {
            \draw[-, thick, draw=violet]  (\mynode.east) --  (\joinednode.west);
        }
        \foreach \mynode/\joinednode in {3-2/3-3, 0-1/0-2, 3-0/3-1}
        {
            \draw[-, thick, draw=violet] (\mynode.north) -- (\joinednode.south);
        }
  
        \foreach \mynode/\joinednode in {2-0/2-1, 1-1/1-2, 2-2/2-3}
        {
            \draw[dashed] (\mynode.north) -- node [right] {1} (\joinednode.south) ;  
        }
        \foreach \mynode/\joinednode in {0-0/0-1, 0-2/0-3, 3-1/3-2}
        {
            \draw[dashed] (\mynode.north) -- node [right] {2} (\joinednode.south) ;  
        }
        \foreach \mynode/\joinednode in {1-0/1-1, 1-2/1-3, 2-1/2-2}
        {
            \draw[dashed] (\mynode.north) -- node [right] {3} (\joinednode.south) ;  
        }
        \foreach \mynode/\joinednode in {3-0/3-1, 3-2/3-3, 0-1/0-2}
        {
            \draw[dashed] (\mynode.north) -- node [right] {4} (\joinednode.south) ;  
        }
        \draw[->-=0.6, thick, draw=violet] (0-3.east) -- (1-3.west);
        \node[] () [left = of 0-3] {\large (a)};    
    \end{tikzpicture}
    \end{minipage}
    \vspace{2em}
    
    \begin{minipage}[t]{0.45\linewidth}
    \begin{tikzpicture}[scale = 0.75, node distance = 0.2cm]
        \foreach \x in {0,...,3}
        \foreach \y in {0,...,3}
        {
            \node[qubit2] (\x-\y) at (\x,\y){};
        }
        \draw[<->, very thick] (0-0.south) to [out=290,in=250] (1-0.south);
        \draw[<->, very thick] (2-0.south) to [out=290,in=250] (3-0.south);
        \draw[-, draw=white] (1-0.south) to [out=290,in=250] node [below] {$U_L$} (2-0.south);
        
        \foreach \mynode/\joinednode in {0-0/1-0, 0-1/1-1, 0-2/1-2, 0-3/1-3, 2-0/3-0, 2-1/3-1, 2-2/3-2, 2-3/3-3}
        {
            \draw[<->, >=latex, thick, draw=teal] (\mynode.center) -- (\joinednode.center);
        }
        \node[] () [left = of 0-3] {\large (b)};
    \end{tikzpicture}
    \end{minipage}
    \begin{minipage}[b]{0.45\linewidth}
    \begin{tikzpicture}[scale = 0.75]
        \foreach \x in {0,...,3}
        \foreach \y in {0,...,3}
        {
            \node[qubit2] (\x-\y) at (\x,\y){};
        }
        \draw[<->, very thick] (1-0.south) to [out=290,in=250] node [below] {$U_R$} (2-0.south);

        \foreach \mynode/\joinednode in {1-0/2-0, 1-1/2-1, 1-2/2-2, 1-3/2-3}
        {
            \draw[<->, >=latex, thick, draw=orange] (\mynode.center) -- (\joinednode.center);
        }
    \end{tikzpicture}
    \end{minipage}
    
    \caption{(a) Vertical hopping term implementation for a $4 \times 4$ grid of fermions. The numbers $i$ show which vertical term will be implemented after $i$ applications of $U_R U_L$. The highlighted blue lines show the only places where the hopping terms can be implemented -- at the JW-adjacent positions. (b) Action of $U_L$ and $U_R$ on the grid of qubits.}
    \label{fig:ansatz_order}
\end{figure}
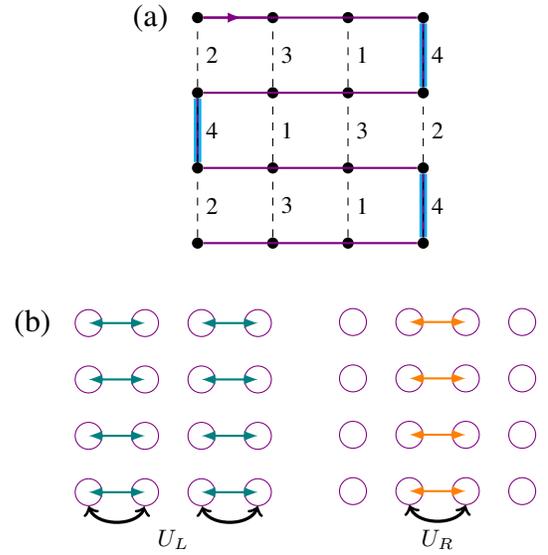

\tikzset{dot/.style={shape=circle,fill=black,scale=0.4}}
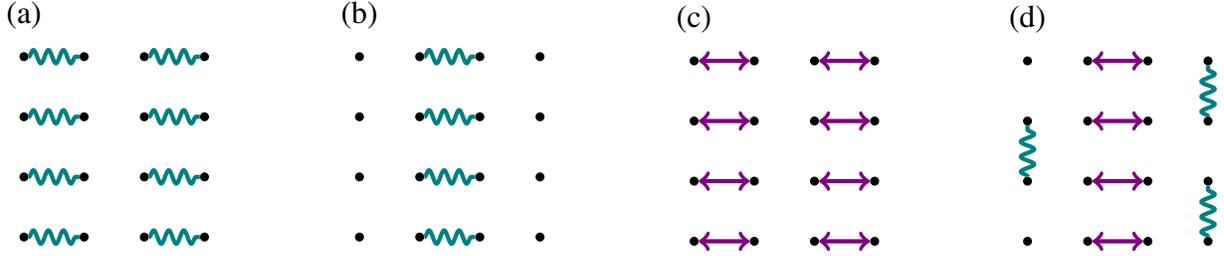
\begin{figure*}[t]
    \centering
    \begin{minipage}[t]{0.2\linewidth}
    \begin{tikzpicture}[node distance = 0.2cm, scale = 0.8]
        \foreach \x in {0,...,3}
        \foreach \y in {0,...,3}
        {
            \node[dot] (\x-\y) at (\x,\y){};
        }
        \foreach \mynode/\joinednode in {0-0/1-0, 0-1/1-1, 0-2/1-2, 0-3/1-3, 2-0/3-0, 2-1/3-1, 2-2/3-2, 2-3/3-3}
        {
            \draw[decorate, decoration={snake, segment length = 0.2cm}, ultra thick, draw=teal] (\mynode.east) -- (\joinednode.west);
        }
        \node[] () [above = of 0-3] {\large (a)};
          \path
    ([shift={(-5\pgflinewidth,-5\pgflinewidth)}]current bounding box.south west)
    ([shift={( 5\pgflinewidth, 5\pgflinewidth)}]current bounding box.north east);
    \end{tikzpicture}
    \end{minipage}
    \hspace{5ex}
    \begin{minipage}[t]{0.2\linewidth}
    \begin{tikzpicture}[node distance = 0.2cm, scale = 0.8]
        \foreach \x in {0,...,3}
        \foreach \y in {0,...,3}
        {
            \node[dot] (\x-\y) at (\x,\y){};
        }
        \foreach \mynode/\joinednode in {1-0/2-0, 1-1/2-1, 1-2/2-2, 1-3/2-3}
        {
            \draw[decorate, decoration={snake, segment length = 0.2cm}, ultra thick, draw=teal] (\mynode.east) -- (\joinednode.west);
        }
        \node[] () [above = of 0-3] {\large (b)};
          \path
    ([shift={(-5\pgflinewidth,-5\pgflinewidth)}]current bounding box.south west)
    ([shift={( 5\pgflinewidth, 5\pgflinewidth)}]current bounding box.north east);
    \end{tikzpicture}
    \end{minipage}
    \hspace{5ex}
    \begin{minipage}[t]{0.2\linewidth}
    \begin{tikzpicture}[node distance = 0.2cm, scale = 0.8]
        \foreach \x in {0,...,3}
        \foreach \y in {0,...,3}
        {
            \node[dot] (\x-\y) at (\x,\y){};
        }
        \foreach \mynode/\joinednode in {0-0/1-0, 0-1/1-1, 0-2/1-2, 0-3/1-3, 2-0/3-0, 2-1/3-1, 2-2/3-2, 2-3/3-3}
        {
            \draw[ultra thick, draw=violet, arrows=<->] (\mynode.east) -- (\joinednode.west);
        }
        \node[] () [above = of 0-3] {\large (c)};
          \path
    ([shift={(-5\pgflinewidth,-5\pgflinewidth)}]current bounding box.south west)
    ([shift={( 5\pgflinewidth, 5\pgflinewidth)}]current bounding box.north east);
    \end{tikzpicture}
    \end{minipage}
    \hspace{5ex}
    \begin{minipage}[t]{0.2\linewidth}
    \begin{tikzpicture}[node distance = 0.2cm, scale = 0.8]
        \foreach \x in {0,...,3}
        \foreach \y in {0,...,3}
        {
            \node[dot] (\x-\y) at (\x,\y){};
        }
        \foreach \mynode/\joinednode in {1-0/2-0, 1-1/2-1, 1-2/2-2, 1-3/2-3}
        {
            \draw[ultra thick, draw=violet, arrows=<->] (\mynode.east) -- (\joinednode.west);
        }
        \foreach \mynode/\joinednode in {3-2/3-3, 0-1/0-2, 3-0/3-1}
        {
            \draw[decorate, decoration={snake, segment length = 0.2cm}, ultra thick, draw=teal] (\mynode.north) -- (\joinednode.south);
        }
        \node[] () [above = of 0-3] {\large (d)};
          \path
    ([shift={(-5\pgflinewidth,-5\pgflinewidth)}]current bounding box.south west)
    ([shift={( 5\pgflinewidth, 5\pgflinewidth)}]current bounding box.north east);
    \end{tikzpicture}
    \end{minipage}
    \caption{
    Quantum circuit elements required to implement one layer of the EHV or NP ansatz for a single spin-type.
    Circuit layers go from (a) to (d), with (c) and (d) repeated thrice more to complete the swap network. Wavy green lines are the number-preserving unitaries $U_{\text{NP}}$. Purple arrows are FSWAP gates, with (c) representing $U_L$ and (d) representing $U_R$ implemented in parallel with vertical hopping terms. In our implementation, the (b) layer is moved to the end, allowing the horizontal hopping terms in (a) and (b) to be combined with the FSWAP gates in (c) and (d) respectively.}
    \label{fig:ansatz_layer}
\end{figure*}

Note that the vertical terms are implemented in a different order to the horizontal terms. If the columns begin in the order $1,2,3,4\dots,n$ (assuming that $n$ is even), then after a single application of $U_RU_L$, they are re-ordered to $2, 4, 1, 6,\dots, n-3, n-1$. Each subsequent application of $U_RU_L$ will place a new even-numbered column to the far left, and a new odd-numbered column to the far right, until $n/2$ applications have seen every even-numbered column at the left, and every odd-numbered column at the right. Since it is at the far ends that vertical terms can be applied locally, then after $n/2$ applications of $U_RU_L$, all terms that can be applied locally at the left will have been applied for the even-numbered columns, and similarly for the odd-numbered columns. Applying another $n/2$ iterations of $U_RU_L$ will see all even-numbered columns move to the right, and all odd-numbered columns to the left, which allows the remaining terms to be implemented locally. Figure \ref{fig:ansatz_order} illustrates the order in which the vertical hopping terms are implemented for a $4 \times 4$ grid of fermions (ignoring spin).

If we assume that gates can be applied across arbitrary pairs of qubits, and that both FSWAP and $U_{\text{NP}}$\footnote{The hopping terms in the HV ansatz $e^{i\theta(XX+YY)/2}$ are a special case of $U_{\text{NP}}$ where $\phi = 0$.} can be implemented in depth 1, then the circuit used to implement all vertical hopping terms will have depth $2n_x$ for even $n_x$, and depth $2n_x+1$ for odd $n_x$. This is because for even $n_x$ the hopping terms can be implemented in parallel with $U_R$, and for odd $n_x$ some hopping terms can be implemented in parallel with $U_L$ and others with $U_R$; one hopping term is left over in the latter case, leading to an overall overhead of 1. All horizontal hopping terms can be implemented in depth 2, and all onsite terms in depth 1. In fact, it is possible to perform a combined horizontal hopping term and FSWAP operation in depth 1\footnote{Up to single qubit gates: FSWAP $\cdot U_{\text{NP}}(\theta,\phi) = (Z^{3/2} \otimes Z^{3/2})\cdot U_{\text{NP}}(\theta + \frac{\pi}{2},\phi)$.}. By replacing the first and last layers of FSWAP gates (the first corresponding to $U_L$, and the last corresponding to $U_R$) with such a combined operation we effectively fold the horizontal hopping terms into the swap network operator $U_RU_L$. Hence, the final depth of the circuit that implements one layer of the ansatz is $2n_x+1$ for even $n_x$, and $2n_x+2$ for odd $n_x$. Figure \ref{fig:ansatz_layer} shows the circuit used to implement a single layer of the ansatz for a $4 \times 4$ grid, for just one of the spins (and therefore omitting the onsite interactions).

We stress that this efficient version of the HV ansatz is different from the standard HV ansatz, in that vertical hopping terms are implemented in a different order. We refer to it as the efficient HV (EHV) ansatz below.

It is worth comparing the complexity of the EHV ansatz to what we would obtain by implementing time-evolution according to each term in the HV ansatz directly. Considering a $2 \times n_y$ grid (the first case where the two ans\"atze differ), using the snake ordering, horizontal hopping terms and onsite interactions can each be implemented in depth 1. Vertical interactions either can be implemented in depth 1, or require an operation of the form $e^{i\theta (XX+YY)ZZ}$ to be implemented. As discussed in Appendix \ref{app:bvc}, this can be achieved with a circuit of depth 4, assuming that arbitrary 2-qubit gates are available. Therefore, the overall depth of the circuit for each layer is $2 + 2 \times (1+4) = 12$, which is more than twice as large. For grids where $n_x$ is larger, the improvement will be even more pronounced.
As another comparison, Reiner et al.~\cite{reiner19} reported a circuit with 81 two-qubit gates per layer for a $3\times 3$ grid, whereas the circuit here would use at most $9 + 2 \times 3 \times (6 + 2) = 57$ two-qubit gates per layer.

Finally, we remark that all this discussion has assumed the use of open boundary conditions in the Hubbard model. Periodic boundary conditions in the horizontal direction can be implemented without any overhead, but periodic boundaries in the vertical direction are significantly more challenging. However, smooth boundary conditions, which can be even more advantageous in terms of reducing finite-size effects~\cite{vekic96}, can also be implemented efficiently.


\subsection{Measurement}
\label{sec:measurement}

At the end of each run of the circuit, we need to measure the energy of the state $\ket{\psi}$ produced with respect to $H$. (Note that, unlike quantum Monte Carlo methods, there is no issue with correlation between runs, and each measurement is assumed to be independent.) The most na\"ive method to achieve this would involve measuring $\braket{\psi|H_i|\psi}$ for each term $H_i$ in $H$. For an $n_x \times n_y$ grid, there are $4n_x n_y - 2n_x - 2n_y$ hopping terms and $n_x n_y$ onsite terms, giving $5n_xn_y -2n_x-2n_y$ terms in total, which can be a significant overhead (e.g.\ 156 terms for $n_x=n_y=6$). Even worse, these terms involve long-range interactions via the Jordan-Wigner transform, suggesting that energy measurement can be challenging.

However, it turns out that many of these terms can be measured in parallel, by grouping them together into at most five commuting sets. There have been a number of recent works on general techniques for splitting the terms of a local Hamiltonian into commuting sets \cite{huggins19, crawford19, gokhale19, Izmaylov2019, gokhale19_2}; here we have a particularly efficient way to do this using the lattice structure of the Hamiltonian. The onsite terms can be measured all at once and the hopping terms can be broken into at most four sets -- two horizontal and two vertical -- as displayed in Figure \ref{fig:hopping_meas}. 

First, the onsite terms can simply be measured by carrying out a computational basis measurement on every qubit. In the Jordan-Wigner picture the onsite terms map to a matrix of the form $\frac{1}{4}(I-Z_i)(I-Z_j) = \ket{11}\bra{11}_{ij}$. So the energy for each term corresponding to a particular site is the probability that the two qubits corresponding to this site (spin up and spin down) are both measured to be in the state 1.

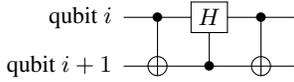
\begin{figure}[t]
    \centering
    \leavevmode
        \Qcircuit @C=1em @R=1em {
        &\lstick{\text{qubit } i} & \ctrl{1} & \gate{H} & \ctrl{1} & \qw\\
        &\lstick{\text{qubit } i+1} & \targ & \ctrl{-1} & \targ & \qw
        }
        \caption{Unitary $U$ that transforms into the $\frac{1}{2}(XX+YY)$ basis.}
        \label{fig:basistransform}
\end{figure}

Horizontal hopping terms take the form $\frac{1}{2}(X_iX_{i+1} + Y_iY_{i+1})$. These terms can be measured efficiently by first transforming into a basis in which this operator is diagonal. This can be done with the quantum circuit $U$ shown in Figure \ref{fig:basistransform}, which diagonalises $\frac{1}{2}(XX+YY)$ as $D = \ket{01}\bra{01} - \ket{10}\bra{10}$ and so the expectation of $\frac{1}{2}(XX+YY)$ is equivalent to the probability of getting the outcome `01' minus the probability of getting `10'. It is important to note that we cannot measure the hopping term on qubit pairs $(i-1,i)$ and $(i,i+1)$ simultaneously due to this transformation, and so if $n_x > 2$ we require two preparations of the ansatz circuit to measure the horizontal hopping terms.

The vertical terms can be measured in a similar way, but with the added complication of the Pauli-$Z$ strings $\frac{1}{2}(X_iX_j + Y_iY_j)Z_{i+1}\cdots Z_{j-1}$. Qubits $i$ and $j$ are treated like the horizontal hopping terms and the $Z$ strings are dealt with by multiplying the expectation by a parity term. Doing a computational basis measurement on qubits $i+1$ to $j-1$ and counting the number of times that `1' is measured gives the parity term. If there are an even number, the parity is 1, otherwise it is $-1$.

All the vertical hopping terms can also be measured with at most two executions of the ansatz circuit. For example, consider the $4 \times 4$ grid shown in Figure \ref{fig:hopping_meas}. For the first set of vertical hopping terms (shown in blue) we can apply $U$ to all eight pairs of qubits corresponding to these terms simultaneously (pairs $(0,7),(1,6),\dots,(11,12))$. Since $U$ has the property that $U^\dagger (Z \otimes Z) U = Z \otimes Z$, we can then collect statistics for measuring $D$ on each pair of qubits and all the required $Z$-strings (e.g.\ $Z_1\dots Z_6$) simultaneously. This is a consequence of the chosen Jordan-Wigner ordering -- there are always an even number of Pauli-$Z$ operators in between qubits $i$ and $j$.

Note that, in our scheme, measurement is the one point in the circuit where quantum gates need to be applied across qubits that are not adjacent with respect to the JW encoding. We also remark that this approach allows a simple notion of error-detection, by checking the Hamming weight of the returned measurement results (see Section \ref{sec:handling_noise}).

Recently, Cai~\cite{cai19} described an alternative approach to obtaining the expectation value using 5 measurements, based on switching the Jordan-Wigner ordering around when measuring the vertical terms, making the vertical hopping terms the JW-adjacent ones and hence removing the Pauli-$Z$ strings. The cost of implementing this approach would be similar to the approach proposed here in the case of square grids (or perhaps slightly more efficient). For non-square grids the approach proposed here will be more efficient, as one can choose the orientation of the grid to minimise the length of Jordan-Wigner strings, whereas the approach of~\cite{cai19} needs to run the quantum circuit twice, one for each orientation.


\subsection{Classical optimiser}

The VQE algorithm makes many calls to the quantum computer to produce trial quantum states. First we will lay out some of the terms that will be important in our analysis.
\begin{itemize}
    \item Circuit evaluation = one run of the quantum computer
    \item Energy measurement = 5 circuit evaluations (see Section \ref{sec:measurement})
    \item Energy estimate = $m$ energy measurements (also referred to as \textit{function evaluation} in the context of optimisation routines)
\end{itemize}

We can determine a rough budget for a reasonable number of calls as follows. We start by assuming that we can perform each 2-qubit quantum gate in 100ns and that measurements are instantaneous (to justify this, even faster gates than this have been demonstrated in superconducting qubit systems~\cite{barends14}, and measurements have been demonstrated that are fast enough that their cost is negligible over the whole circuit~\cite{walter17}). Assume for simplicity that the depth of the whole circuit is 100, and that the cost of classical computation is negligible. Then $10^5$ runs of the quantum computer can be executed per second. If we would like to ultimately estimate the energy up to an accuracy of $\sim 10^{-2}$, approximately $10^4$ circuit evaluations are required to estimate each of the 5 terms (see Figure \ref{fig:statistical_error} in the Appendix for numerical results to justify this assumption, where for a particular instance, we found that $m$ measurements achieved energy error $\approx 1.3/\sqrt{m}$). Thus approximately 2 energy estimates up to this precision can be obtained per second. So in $5 \times 10^4$ seconds, corresponding to approximately 14 hours, we can produce approximately $10^5$ energy estimates up to an accuracy of $\sim 10^{-2}$. This motivates us to use $\sim 10^5$ as the budget for the number of function evaluations used by the optimiser. (In fact, in our numerical experiments below, we found that substantially fewer evaluations were sufficient.)

We evaluated different optimisation methods given in the NLopt C library for nonlinear optimisation~\cite{nlopt} and found that L-BFGS was usually a very effective algorithm to use when considering a perfect, noiseless, version of VQE with simulated exact energy measurements. Other algorithms required many more iterations, or often found lower-quality local minima. To estimate the gradient, as required for L-BFGS, we used a simple finite difference approximation.

Including realistic measurements turns the optimisation problem into a stochastic one. In this setting we found that standard deterministic optimisation methods provided by NLopt were ineffective (either failing completely, or producing low-quality results). We therefore turned to stochastic optimisation methods such as the SPSA algorithm~\cite{spsa}, which has been successfully used in VQE experiments on superconducting hardware~\cite{kandala17, Ganzhorn2019}, and a coordinate descent algorithm~\cite{nakanishi19,ostaszewski19,parrish19} that has been shown to be effective for small VQE instances. We remark that, during preparation of this work, alternative stochastic optimisation techniques for VQE have been developed~\cite{kubler19,wilson19,Verdon2019}; evaluating and improving such techniques in the context of the Hubbard model is an important direction for future work.


\subsubsection{Simultaneous perturbation stochastic approximation}
\label{sec:spsa}
The simultaneous perturbation stochastic approximation (SPSA) algorithm~\cite{spsa} works in a similar way to the standard gradient descent algorithm, but rather than estimating the full gradient, instead picks a random direction to estimate the gradient along. This is intended to make SPSA robust against noise and to require fewer function evaluations. Many aspects of this algorithm can be tailored to the specific problem at hand, such as parameters that govern the rate of convergence, terminating tolerances and variables on which the tolerance is monitored, and the number of gradient evaluations to average the estimated gradient over.

Each gradient evaluation is estimated from two function evaluations (as compared with typically twice the number of parameters for finite difference methods) and is given by
\begin{equation*}
    g(\bm{\theta}_k) = \frac{f(\bm{\theta}_k + c_k\bm{\Delta}_k) - f(\bm{\theta}_k - c_k\bm{\Delta}_k)}{2 c_k} \bm{\Delta}_k^{-1}, 
\end{equation*}
where $\bm{\theta}_k$ is the current parameter vector after $k$ steps, $c_k$ is an optimisation parameter to be determined, and the parameters are perturbed with respect to a Bernoulli $\pm 1$ distribution $\bm{\Delta}_k$ with probability $\frac{1}{2}$ for each outcome. The gradient step size is $c_k = c/(k+1)^\gamma$, where in our experiments $\gamma = 0.101$ was chosen to be the ideal theoretical value~\cite{Spall1998} and $c = 0.2$. The parameters are then updated via
\begin{equation*}
    \bm{\theta}_{k+1} = \bm{\theta}_k - a_k g(\bm{\theta}_k)
\end{equation*}
where $a_k = a/(k+1+A)^\alpha$ dictates the speed of convergence. Similarly to $\gamma$, $\alpha = 0.602$ is chosen as the ideal theoretical value \cite{Spall1998}, while we set the stability constant $A = 100$ and $a = 0.15$. The values of $a$ and $c$ were chosen by a joint parameter sweep. We found that the parameters generally had to be small to reduce the rate of convergence, which allowed us to reach a more accurate result but with more iterations.

\begin{figure}[t]
    \centering
    \input{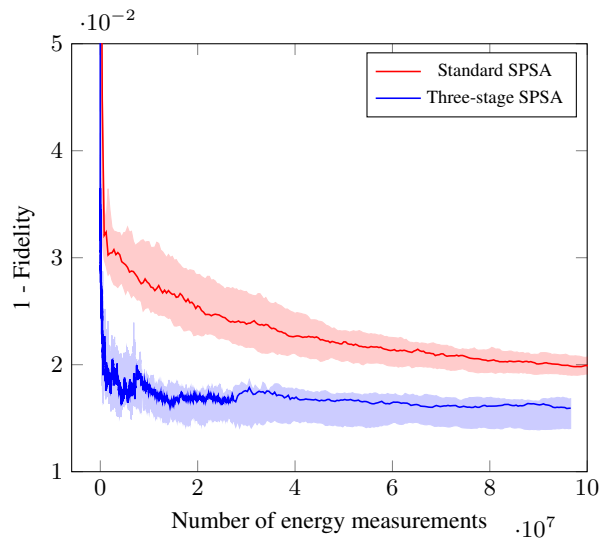}
    \caption{Infidelity achieved over 5 runs of the standard SPSA algorithm (where each energy estimate is formed of $10^4$ energy measurements and two gradient evaluations are taken in each iteration) and a modified three-stage SPSA algorithm which starts with less accurate measurements, as described in the text. Results are shown for a $1\times 6$ grid, EHV ansatz, depth 5. The solid lines show the median of the runs and the limits of the shaded regions are the maximum and minimum values seen over the 5 runs.}
    \label{fig:spsa_rough_smooth}
\end{figure}

The main modification we made to the standard SPSA algorithm is to perform multiple runs of the optimiser. We start with two coarse runs with a high level of statistical noise where we calculate the energy estimate using only $10^2$ and then $10^3$ energy measurements. This is followed by a finer run where SPSA is restarted using $10^4$ energy measurements for the estimate and averaging over two gradient evaluations in random directions for $g(.)$. The number of steps in this three stage optimisation is determined by a ratio of $10:3:1$. 
Figure \ref{fig:spsa_rough_smooth} shows the beneficial effect of starting by making less accurate measurements, as described.

\subsubsection{Coordinate descent algorithm}

We now describe an alternative algorithm, based on an approach independently discovered by~\cite{nakanishi19,ostaszewski19,parrish19}. The basic algorithm presented in these works can be applied to variational ans\"atze where the gates are of the form $e^{i\theta H}$ for Hamiltonians $H$ such that $H^2=I$ (e.g.\ Pauli matrices). It is based on the nice observation that, for gates of this form, the energy of the corresponding output state is a simple trigonometric polynomial in $\theta$ (if all other variational parameters are fixed). This implies that it is sufficient to evaluate the energy at a small number (three) of choices for $\theta$ in order to {\em analytically} determine its minimum with respect to $\theta$. The algorithm proceeds by choosing parameters in some order (e.g.\ a simple cyclic sequential order, or randomly) and minimising with respect to each parameter in turn. It is shown in~\cite{nakanishi19,ostaszewski19,parrish19} that this approach can be very effective for small VQE instances.

We use a generalisation of this approach which works for any Hamiltonian with integer eigenvalues. This enables us to apply the algorithm to the number-preserving (and hence HV) ansatz, because each gate in the ansatz can be seen as combining the pair of gates $e^{i\theta(XX+YY)/2}$, $e^{i\phi \proj{11}}$. The corresponding Hamiltonians have eigenvalues $\{0,\pm1\}$, $\{0,1\}$ respectively. The generalisation is effectively the same as the one presented in~\cite{nakanishi19,parrish19} to optimise over separate gates which share the same parameters. However, here we present the algorithm and its proof somewhat differently and include a full argument for how to compute the minimum with respect to $\theta$, which is not included in~\cite{nakanishi19,parrish19}.

The algorithmic approach has been given different names in the literature (``sequential minimal optimization''~\cite{nakanishi19}, ``Rotosolve''~\cite{ostaszewski19}, ``Jacobi diagonalization''~\cite{parrish19}). Here we prefer yet another name, coordinate descent~\cite{tseng01} (CD), because this encompasses the approach we consider, whereas the above names technically refer to special cases of the approach which are not directly relevant to the algorithm we use\footnote{Sometimes the term ``coordinate descent'' is used for algorithms that perform gradient descent in each coordinate; we stress that here we instead {\em exactly} minimise over each coordinate.}.

Let $A$ be a Hermitian matrix with eigenvalues $\lambda_k \in \Z$, and assume that $e^{i\theta A}$ is one of the gates (parametrised by $\theta$) in a variational ansatz. Then the energy of the output state with respect to $H$ can be written as
\[ \tr[H U e^{i \theta A} \proj{\psi} e^{-i\theta A} U^\dag] \]
for some state $\ket{\psi}$ and unitary operator $U$ that do not depend on $\theta$. Writing $A = \sum_k \lambda_k P_k$ for some orthogonal projectors $P_k$ and using linearity of the trace, this expands to
\[ \sum_{k,l} e^{i \theta(\lambda_k-\lambda_l)} \tr[H U P_k \proj{\psi} P_l U^\dag]. \]
If $\Delta$ denotes the set of possible differences $\lambda_k - \lambda_l$, and $D = \max_{k,l} |\lambda_k-\lambda_l|$, this expression can be rewritten as
\[ f(\theta) = \sum_{\delta \in \Delta} c_\delta e^{i \theta \delta} \]
for some coefficients $c_\delta \in \C$. This is a (complex) trigonometric polynomial in $\theta$ of degree $D$. So it can be determined completely by evaluating it at $2D+1$ points. A particularly elegant choice for these is $\theta \in \{2k\pi/(2D+1) : -D \le k \le D\}$. Then the coefficients $c_k$ can be determined via the discrete Fourier transform:
\[ c_k = \frac{1}{2D+1} \sum_{l=-D}^D e^{-2\pi i kl / (2D+1)} f(2\pi l / (2D+1)). \]
To minimise $f$, we start by computing the derivative
\[ \frac{df}{d\theta} = i \sum_{k=-D}^D k c_k e^{ik\theta}, \]
and finding the roots of this function. To find these roots, we consider the function $g(\theta) = e^{2iD\theta} \frac{df}{d\theta}$. Every root of $\frac{df}{d\theta}$ is a root of $g(\theta)$, and as $g(\theta)$ is a polynomial of degree $2D$ in $e^{i\theta}$, its roots can be determined efficiently (e.g.\ by computing the eigenvalues of the companion matrix of $g$).

Finally, we ignore all roots that do not have modulus 1 (i.e.\ consider only roots of the form $e^{i\theta}$) and choose the root $e^{i\theta_{\min}}$ at which $f(\theta_{\min})$ is smallest. Note that the only steps throughout this algorithm which require evaluation of $f(\theta)$ using the quantum hardware are the $2D+1$ evaluations required for polynomial interpolation.

The above argument extends to the situation where we have $m$ Hamiltonian evolution operations in the circuit that all depend on the same parameter $\theta$; in this case, one obtains a trigonometric polynomial of degree $mD$ (see~\cite{nakanishi19,parrish19} for a proof), which is determined by its values at $2mD + 1$ points. This enables us to apply this optimisation algorithm to the (efficient) Hamiltonian Variational ansatz as well.



\subsection{Handling noise}
\label{sec:handling_noise}

The VQE approach needs to contend with two different kinds of noise: statistical noise inherent to the quantum measurement process, and errors in the circuit. Statistical noise can be mitigated by simply taking more measurements, while the ans\"atze we use allow for a simple notion of error-detection with no overhead in terms of number of qubits or execution time. The NP (and hence HV) ansatz corresponds to quantum circuits where every operation in the circuit preserves fermionic occupation number (equivalently, Hamming weight after the Jordan-Wigner transform). So, if the final state of the quantum algorithm contains support on computational basis states of different Hamming weight to the start of the algorithm, one can be confident that an error has occurred.

Further, the Hamming weight of the final state can be measured as part of the measurement procedure described in Section \ref{sec:measurement} without any additional cost. Onsite energy measurements simply correspond to measurements in the computational basis, while measurements corresponding to hopping terms split pairs of qubits according to the pair's total Hamming weight. So Hamming weights of pairs (and hence the total Hamming weight) can be determined simultaneously with measuring according to hopping terms.


\section{Numerical validation}

We developed a high-performance software tool in C++, based on the Quantum Exact Simulation Toolkit~\cite{jones19} (QuEST), which enabled the ans\"atze we used to be validated and compared.
The tests were mainly carried out on the Google Cloud Platform.
In the preliminary tests, we found that GPU-accelerated QuEST commonly outperformed QuEST running on CPU only (whether single-threaded, multi-threaded, or distributed).
For most of the results reported here, we found a speed-up of 4-5x when compared with a 16 vCPU machine (n1-highcpu-16) available on Google Cloud, which is similar to the speed-up reported in~\cite{jones19}.
The GPU-accelerated tests were carried out using a single vCPU machine (n1-standard-1) equipped with either NVIDIA Tesla P4 (nvidia-tesla-p4) or NVIDIA Tesla K80 (nvidia-tesla-k80).
Some of the noisy experiments were carried out on a single vCPU instance (n1-standard-1), as for some of the smaller grid sizes it was found that a single CPU performs similarly to a GPU-accelerated version (for small grid sizes, the data transfer between CPU and GPU dominates the run-time).

We carried out the following tests. First, we tested the expressivity of the HV, efficient HV (``EHV''), and NP ans\"atze by simulating the VQE algorithm using these ans\"atze, with (unrealistic) exact energy measurements, and increasing circuit depths and grid sizes. This builds confidence that the variational approach will be effective for grid sizes beyond those that can be simulated with classical hardware.
Next, we tested the effect of realistic energy measurements; that is, we simulate the entire variational process, including measuring the energy via the procedure described in Section \ref{sec:measurement}. Finally, we tested the effect of noise in the quantum circuit. By contrast with the coherent errors considered in~\cite{reiner19}, we used a depolarising noise model.

For realistic energy measurements we obtained a significant speedup by storing the probability amplitudes of the final state produced by the circuit. Computational basis measurements on that state were then simulated by sampling from this distribution, hence avoiding the need to rerun the circuit. This optimisation is not available with noisy circuits, so those tests are much more computationally intensive.

We now outline some implementation decisions that were made. First, unless specified otherwise, we started with the number of occupied orbitals that corresponds to the lowest energy of the Hamiltonian $H$ defined in (\ref{eq:hubbard}) (not e.g.\ the half-filled case as in~\cite{wecker15}). These occupation numbers are listed in Table \ref{tab:occ_orb}. The ans\"atze we use preserve fermion number, so remain in this subspace throughout the optimisation process.

\begin{table}[t]
\begin{tabular}{|c|c|}
\hline
Occupied orbitals & Grid sizes \\ \hline
2 & $1 \times 2,\, 1 \times 3,\, 2 \times 2$ \\
3 & $1 \times 4$ \\
4 & $1 \times 5,\, 1 \times 6,\, 2 \times 3$ \\
6 & $1 \times 7,\, 1 \times 8,\, 2 \times 4,\, 3 \times 3$ \\
7 & $1 \times 9$ \\
8 & $1 \times 10,\, 1 \times 11,\, 2 \times 5,\, 2 \times 6$ \\
9 & $1 \times 12,\, 3 \times 4$ \\\hline
\end{tabular}
\caption{Number of occupied orbitals corresponding to the lowest energy of the Hubbard Hamiltonian for each grid size tested.}
\label{tab:occ_orb}
\end{table}

For the HV ansatz, one needs to choose the ordering of Hamiltonian terms for time-evolution (see (\ref{eq:hv}). By contrast, for the two ``efficient'' ans\"atze, this ordering is largely pre-determined, except that we have a choice of when to implement the onsite terms in the EHV ansatz; we chose to do so at the start of each layer. In the case of $1\times n_y$ grids, we used a $O,V_1,V_2$ ordering. For $2 \times n_y$ grids, we used a $O,H,V_1,V_2$ ordering (except $2 \times 2$, where there is no $V_2$ term). For $3 \times n_y$ grids, we used an $O,H_1,V_1,V_2,H_2$ ordering.

For all ans\"atze, one needs to choose initial parameters. We used a simple deterministic choice of initial parameters, which (similarly to~\cite{reiner19}) were all set to $1/L$, where $L$ is the number of layers. We also experimented with choosing initial parameters at random, e.g.\ within the range $[0,2\pi/100]$; this achieved similar performance, suggesting that the optimisation does not experience significant difficulties with local minima. In all cases, the initial state was the ground state of the non-interacting model (see Appendix \ref{app:np} for a discussion of the effect of starting in a computational basis state).


\subsection{Ability to represent ground state of the Hubbard model}

\begin{figure}[t]
    \centering
    \begin{tikzpicture}
    \begin{axis}[xlabel=Grid height $n$,ylabel=Depth to 0.99 fidelity, legend style = {font=\scriptsize, at = {(0.5,1.03)}, anchor=south, legend columns={3}, /tikz/every even column/.append style={column sep=0.35cm}},  yminorgrids=true, ymajorgrids=true, minor y tick num=4, major grid style={thick}]
    \addplot[color=red, mark=square*, mark size=1pt, thick] coordinates {(2, 1) (3, 3) (4, 3) (5, 5) (6, 5) (7, 7) (8, 7) (9, 7) (10, 9) (11, 10) (12, 9)};
    \addlegendentry{$1\times n$ HV};
    \addplot[color=red, mark=*, mark size=1pt, thick, dashed, mark options=solid] coordinates {(2, 1) (3, 3) (4, 3) (5, 5) (6, 5) (7, 7) (8, 7) (9, 7) (10, 9) (11, 10) (12, 9)};
    \addlegendentry{$1\times n$ EHV};
    \addplot[color=red, mark=x, dotted, mark size=2pt, mark options=solid, thick] coordinates {(2, 1) (3, 2) (4, 2) (5, 3) (6, 3) (7, 4) (8, 5) (9, 4) (10, 5) (11, 6)};
    \addlegendentry{$1 \times n$ NP}
    \addplot[color=teal,  mark=square*, mark size=1pt, thick] coordinates {(2, 1) (3, 4) (4, 8) (5, 8) (6, 10)};
    \addlegendentry{$2\times n$ HV};
    \addplot[color=teal, mark=*, mark size=1pt, thick, dashed, mark options=solid] coordinates {(2, 1) (3, 3) (4, 11) (5, 11) (6, 18)};
    \addlegendentry{$2\times n$ EHV};
    \addplot[color=teal, mark=x, mark size=2pt, thick, dotted, mark options=solid] coordinates {(2, 1) (3, 4) (4, 7) (5, 6)};
    \addlegendentry{$2\times n$ NP};
    \addplot[color=blue,  mark=square*, mark size=1pt, thick] coordinates {(3, 4) (4, 10)};
    \addlegendentry{$3\times n$ HV};
    \addplot[color=blue, mark=*, mark size=1pt, thick, dashed, mark options=solid] coordinates {(3, 6) (4, 14)};
    \addlegendentry{$3\times n$ EHV};
    \addplot[color=blue, mark=x, mark size=2pt, thick, dotted, mark options=solid] coordinates {(3, 5)};
    \addlegendentry{$3\times n$ NP};
    \end{axis}
    \end{tikzpicture}
\caption{Depths required (in terms of ansatz layers) to represent the ground state of the $n_x \times n_y$ Hubbard model for the HV, EHV and NP ans\"atze. Each point corresponds to the minimal-depth circuit instance we found (using the L-BFGS optimiser) that produces a final state with fidelity at least $0.99$ with the true Hubbard model ground state ($t=1$, $U=2$). Tests run for all grids of size $n_x n_y \le 12$. For $1\times n$ grids, HV and EHV are the same.}
\label{fig:repdepths}
\end{figure}
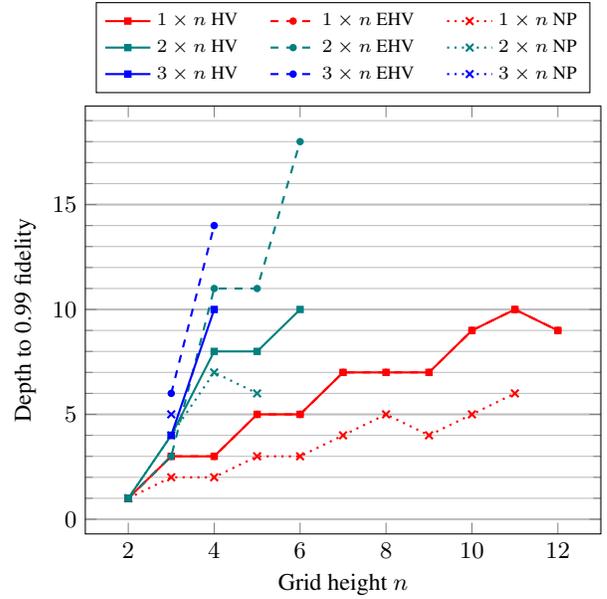

The circuit ans\"atze we consider are divided into layers, and as the number of layers increases, the representational power of the ansatz increases. An initial test of the power of the variational method for producing ground states of the Hubbard model is to determine the number of layers required to produce the ground state $\ket{\psi_G}$ to fidelity $0.99$ where
\begin{equation*}
    \text{Fidelity}(\ket{\psi}) = |\langle \psi_G | \psi \rangle |^2.
\end{equation*}
In Figure \ref{fig:repdepths} we show this for the HV, EHV, and NP ans\"atze. This illustrates that the EHV ansatz (which can be implemented efficiently) performs relatively well in comparison with the well-studied HV ansatz. In most cases (except the $2\times 3$ grid), the HV ansatz requires a lower number of layers, but this is outweighed by the depth reduction per layer achieved by using the EHV ansatz. Note that in the case $n_x=1$, the two ans\"atze are equivalent.

Figure \ref{fig:repdepths} also illustrates that the NP ansatz generally requires lower depth than the other two ans\"atze to achieve high fidelity. This is expected, as it corresponds to optimising over a larger set of circuits. However, it illustrates that the optimisation procedure does not experience any significant difficulties with this larger set other than increased runtime, corresponding to the larger number of parameters. This increase can be significant; e.g.\ a $1\times 11$ grid required approximately $10^5$ function evaluations and a runtime of 16.5 hours on a GPU-accelerated system to achieve fidelity 0.99 using the NP ansatz, whereas achieving the same fidelity using the EHV ansatz required fewer than 9000 function evaluations and a runtime of 1.5 hours.

\begin{figure}[t]
    \centering
    \begin{tikzpicture}
    \pgfplotstableread{diagrams/results/depthscaling.txt} \datatable
    \begin{semilogyaxis}[xlabel=Ansatz depth,ylabel=1 - Fidelity,legend style={font=\scriptsize, at = {(0.96,0.95)}}, ymode=log, grid=both]
        \addplot table[x=depth, y=1x12] from \datatable ;
        \addlegendentry{$1 \times 12$}
        \addplot table[x=depth, y=2x6] from \datatable ;
        \addlegendentry{$2 \times 6$}
        \addplot[color=teal, mark=triangle*, mark size=3pt] table[x=depth, y=3x4] from \datatable;
        \addlegendentry{$3 \times 4$}
    \end{semilogyaxis}
    \end{tikzpicture}
    
    \caption{Scaling of infidelity ($1-$fidelity) with number of layers of EHV ansatz for grids with 12 sites.}
    \label{fig:depthscaling}
\end{figure}
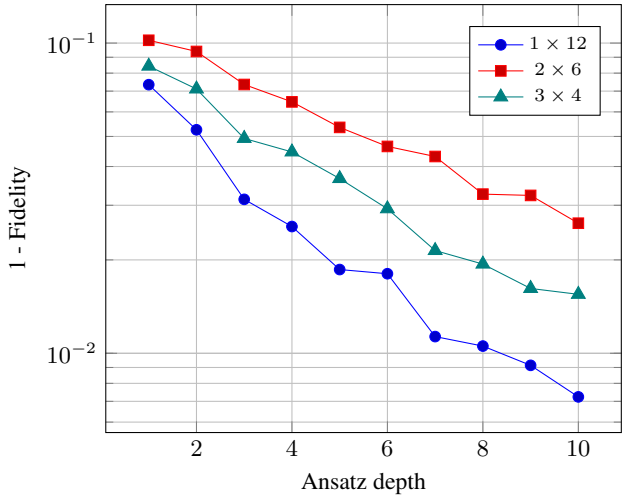

In Figure \ref{fig:depthscaling} we illustrate how the fidelity improves with depth using the EHV ansatz, for the largest grid sizes we considered. In each case, the infidelity decreases exponentially with depth. Notably, $2 \times 6$ seems to be more challenging than $3\times 4$.

\subsection{Optimisation with realistic measurements}

We compared the ability of the SPSA and CD algorithms to find the ground state of Hubbard model instances for four representative grid sizes: $2\times2$, $1\times6$, $2\times3$, and $3\times3$.
For CD, we fixed the number of approximate energy estimates to $\sim 1.2 \times 10^3$, where each estimate consists of $10^4$ energy measurements.
This translates to a limit of $\sim 6 \times 10^7$ circuit evaluations.
For SPSA, on the other hand, the number of energy estimates was limited to $\sim 1.2 \times 10^4$, due to the number of measurements per estimate changing throughout the course of the optimization.
As described in Section \ref{sec:spsa}, we carry out a three-stage optimisation routine and set the ratio of $10:3:1$ for very coarse, coarse, and smooth function evaluations, respectively.
By limiting to a total of $\sim 1.2 \times 10^4$ energy estimates, we allow for a similar total limit as that of CD, $\sim 1.2 \times 10^7$ energy measurements (or $\sim 6 \times 10^7$ circuit evaluations).

For each grid size, we determined the final fidelity of the output of the VQE algorithm with the true ground state after the fixed number of measurements.
For the circuit depth, we chose the minimal depth for which the ground state is achievable (via Figure \ref{fig:repdepths}).

\begin{figure}[t]
    \centering
    \input{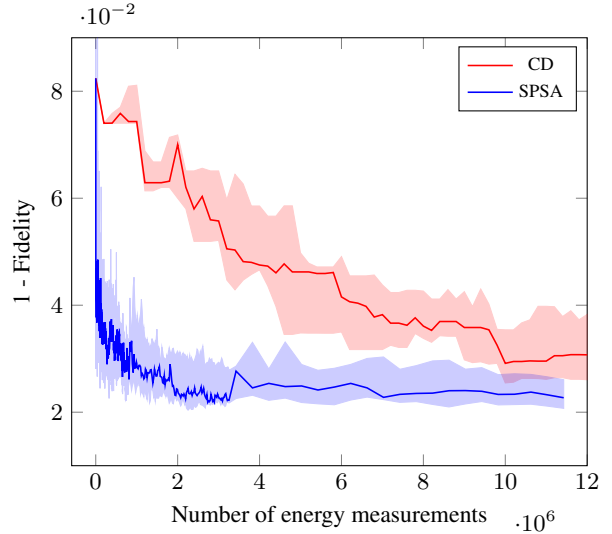}
    \caption{Infidelity reached during the optimisation process with CD and SPSA optimisers and realistic measurements. Results are shown for 5 runs of a $3\times 3$ grid, EHV ansatz, depth 6. The solid lines show the median of the runs and the limits of the shaded regions are the maximum and minimum values seen over the 5 runs.}
    \label{fig:realistic}
\end{figure}

The results are shown in Figure \ref{fig:realistic} and Table \ref{tab:realisticfidelities}. In all cases, both algorithms are able to achieve relatively high fidelity (considering that each energy measurement involves at most $10^4$ circuit runs, suggesting an error of $\sim 10^{-2}$). However, in the case of $1\times 6$ and $3\times 3$ grids, SPSA achieves a noticeably higher fidelity. It is also interesting to note in Figure \ref{fig:realistic} that SPSA uses substantially fewer energy measurements to achieve a high fidelity. One reason for this may be that each iteration of CD requires more energy measurements ($2 n_x n_y + 1 = 19$ for a $3\times 3$ grid, as compared with 2 energy measurements for SPSA).

\begin{table}[t]
    \centering
    \begin{tabular}{|c|c|c|c|c|}
        \hline Grid & Depth & CD & SPSA & L-BFGS\\
         \hline $2\times 2$ & 1 & 0.0068 & 0.0066 & 0.0066 \\
         $1\times 6$ & 5 & 0.0293 & 0.0199 & 0.0098 \\
         $2\times 3$ & 3 & 0.0202 & 0.0199 & 0.0075 \\
         $3\times 3$ & 6 & 0.0307 & 0.0227 & 0.0068 \\
         \hline
    \end{tabular}
    \caption{Final infidelity reached for CD and SPSA optimisers and realistic measurements, compared with the best infidelity achieved by the L-BFGS optimiser with exact measurements. EHV ansatz. CD and SPSA results are median of 5 runs.}
    \label{tab:realisticfidelities}
\end{table}

\subsection{Optimisation with noisy quantum circuits}

We next evaluated the effect of noise on the ability of the VQE algorithm to find the ground state of the Hubbard model. We considered a simple depolarising noise model where, after each 2-qubit gate, each qubit experiences noise with probability $p$ (modelled as Pauli $X$, $Y$, $Z$ operations occurring with equal probability). We examined noise rates $p \in \{10^{-3},10^{-4},10^{-6}\}$ and grid sizes $2 \times 2$, $1 \times 6$ and $2 \times 3$. These experiments are substantially more computationally costly than those with realistic measurements.

We tested the effect of the error-detection procedure described in Section \ref{sec:handling_noise}. When an error is detected by the Hamming weight being incorrect, that run is ignored, and the measurement procedure continues until the intended number of valid energy measurements are produced for each type of term. Hence the total number of energy measurements is somewhat larger than the noiseless case.

\begin{figure}[t]
    \centering
    \input{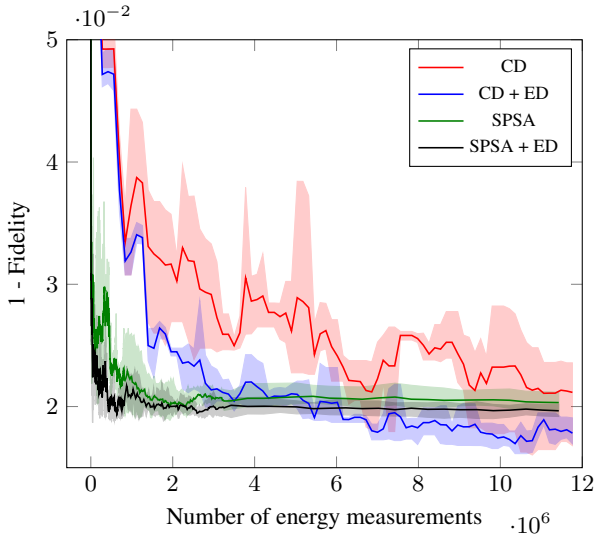}
    \caption{Infidelity reached during the optimisation process with CD and SPSA optimisers, with and without error detection (ED). $2\times 3$ grid, $10^{-3}$ error rate, EHV ansatz, depth 3. The solid lines show the median of the runs and the limits of the shaded regions are the maximum and minimum values seen over the 3 runs.}
    \label{fig:noisy2x3}
\end{figure}

We list the final fidelities achieved for different grid sizes, error rates, and optimisation algorithms in Table \ref{tab:noisyresults}. An illustrative set of runs for a $2\times 3$ grid is shown in Figure \ref{fig:noisy2x3}. The overhead of error-detection is not shown in this figure (that is, measurements where an error is detected are not counted). One can see that in all cases, errors do not make a significant difference to the final fidelities achieved, compared with the noiseless results in Table \ref{tab:realisticfidelities}. The use of error-detection seems to usually lead to a small but noticeable improvement in the final fidelity achieved, as well as seeming to make the performance of the optimiser during a run less erratic. We note that error detection might have a more relevant role for bigger grid sizes, due to higher depths and longer circuit run times. However, more detailed experiments would be required to fully assess the benefit of error-detection. 

\begin{table*}[t]
    \centering
    \begin{tabular}{|c|c|c|c|c|}
    \hline $2\times2$ & \multicolumn{2}{c|}{CD} & \multicolumn{2}{c|}{SPSA} \\
    \hline & No ED & ED & No ED & ED \\
    \hline $10^{-3}$ & 0.0066 & 0.0066 & 0.0066 & 0.0067 \\
    $10^{-4}$ & 0.0067 & 0.0068 & 0.0066 & 0.0066 \\
    $10^{-6}$ & 0.0065 & 0.0064 & 0.0066 & 0.0066 \\
    \hline
    \end{tabular}
    \hspace{1cm}
    \begin{tabular}{|c|c|c|c|c|}
    \hline $1\times 6$ & \multicolumn{2}{c|}{CD} & \multicolumn{2}{c|}{SPSA} \\
    \hline & No ED & ED & No ED & ED \\
    \hline $10^{-3}$ & 0.0262 & 0.0297 & 0.0187 & 0.0176 \\
    $10^{-4}$ & 0.0250 & 0.0259 & 0.0188 & 0.0180 \\
    $10^{-6}$ & 0.0288 & 0.0257 & 0.0197 & 0.0185 \\
    \hline
    \end{tabular}
    \hspace{1cm}
    \begin{tabular}{|c|c|c|c|c|}
    \hline $2\times3$ & \multicolumn{2}{c|}{CD} & \multicolumn{2}{c|}{SPSA} \\
    \hline & No ED & ED & No ED & ED \\
    \hline $10^{-3}$ & 0.0231 & 0.0174 & 0.0201 & 0.0196 \\
    $10^{-4}$ & 0.0169 & 0.0179 & 0.0194 & 0.0195\\
    $10^{-6}$ & 0.0174 & 0.0183 & 0.0199 & 0.0194\\
    \hline
    \end{tabular}
    \caption{Infidelities at end of runs for varying grid sizes and noise rates, with error detection off/on. Median of 3 runs. EHV ansatz, depths 1, 5, 3 respectively.}
    \label{tab:noisyresults}
\end{table*}

\section{Concluding remarks}

We have carried out a detailed study of the complexity of variational quantum algorithms for finding the ground state of the Hubbard model. Our numerical results are consistent with the heuristic that the ground state of an instance on $N$ sites could be approximately produced by a variational quantum circuit with $\sim N$ layers (and in all cases we considered, the number of layers required was at most $1.5N$).

If only around $N$ layers are required, then the ground state of a $5 \times 5$ instance (larger than the largest instance solved classically via exact diagonalisation~\cite{yamada05}) could be found using a quantum circuit on 50 qubits with around 25 layers, corresponding to an approximate two-qubit gate depth of $24 + 25\times(2\times 5 + 2) + 1 = 325$ in a fully-connected architecture, including the depth required to produce the initial state. This is significantly lower than the complexity for time-dynamics simulation reported in~\cite{childs18,babbush18,nam19}, but is still beyond the capabilities of today's quantum computing hardware. Although our work considered only relatively shallow quantum circuit depths, the ability of the NP ansatz to find ground states suggests that the classical optimisation routines used could continue to work for these deeper circuits, as this ansatz used a much larger number of parameters, e.g.\ over 400 for the largest grids we considered.

While the Hubbard model is an important benchmark system in its own right, its simple structure facilitates an easier implementation of VQE than for typical electronic structure Hamiltonians. An important direction for future work is to carry out a similarly detailed analysis of the complexity of VQE for other practically relevant electronic systems.

Determining the optimal choice of classical optimiser remains an important challenge. It is plausible that the optimisers used here could be combined or modified to improve their performance, and other methods that have been studied contemporaneously with this work include adaptive optimisation algorithms~\cite{kubler19} and techniques based on machine learning or ``meta-learning''~\cite{Verdon2019,wilson19}. Future work should evaluate such methods for larger-scale instances of the Hubbard model and other challenging problems in many-body physics.

{\bf Note.} While finalising this paper, we became aware of a related recent work~\cite{cai19} which also determines theoretical resource estimates for applying the HV ansatz to solve the Hubbard model via VQE. The results obtained are qualitatively similar to ours; our circuit complexity bounds are lower, although the gate count estimates of~\cite{cai19} use a more restrictive gate set and topology targeted at efficient implementation on a specific hardware platform, so are not directly comparable. For example, if solving a $5 \times 5$ instance with a 10-layer HV ansatz, Ref.~\cite{cai19} would estimate a complexity of 11,300 2-qubit gates. By contrast, our estimate with unrestricted 2-qubit gates and interaction topology (see  (\ref{eq:gatecomplexityodd})) is fewer than 3,351 2-qubit gates. The implementation strategy of~\cite{cai19} uses only nearest-neighbour interactions; the strategy discussed in Section~\ref{sec:implementation} for a nearest-neighbour architecture is similar, but with some small differences.


\subsection*{Acknowledgements}

Data are available at the University of Bristol data repository, data.bris, at \url{https://doi.org/10.5523/bris.1873owc1bcmrw1y4raeeuygzuy}. We would like to thank Toby Cubitt, John Morton, and the rest of the Phasecraft team for helpful discussions and feedback, and Zhenyu Cai and Craig Gidney for comments on a previous version. LM received funding from the Bristol Quantum Engineering Centre for Doctoral Training, EPSRC Grant No. EP/L015730/1. Google Cloud credits were provided by Google via the EPSRC Prosperity Partnership in Quantum Software for Modeling and Simulation (EP/S005021/1).

\appendix

\section{Alternative fermion encodings}
\label{app:otherencodings}

A well-known issue with the Jordan-Wigner transform is that it can produce qubit Hamiltonians which contain long strings of $Z$ operators, leading to high-depth quantum circuits. This prompts us to consider alternative encodings of fermions as qubits which could reduce this depth. Here we evaluate two prominent encodings which produce qubit operators whose locality does not depend on the size of the grid. Although we have not proven that the time-evolution circuits we find are optimal, they provide an indication of the relative complexity of these encodings.


\subsection{Ball-Verstraete-Cirac encoding}
\label{app:bvc}

The encoding which (for arbitrary-sized grids) produces the lowest-weight qubit operators known is the Ball-Verstraete-Cirac or auxiliary fermion encoding~\cite{whitfield16}, developed independently in~\cite{ball05,verstraete05}.

The Ball-Verstraete-Cirac encoding can be seen as an optimised Jordan-Wigner encoding that avoids the need for long $Z$ strings, at the expense of adding more qubits. Each fermionic mode (with the possible exception of two of the corners of the grid) is associated with an auxiliary mode, and vertical hopping terms use these modes. In this section, we change notation slightly and let operators of the form $X_{k,l}$ denote Pauli operators acting on the site $k,l$, while letting primed operators of the form $X'_{k,l}$ denote Pauli operators acting on the auxiliary mode associated with site $k,l$. Although there is some freedom in the encoding, the simplest mapping of the hopping terms presented in~\cite{verstraete05} is as follows. Each vertical hopping term $a^\dag_{k,l} a_{k,l+1} + a^\dag_{k,l+1} a_{k,l}$ maps to either
\[ V_{k,l} := (-1)^{l+1} (X_{k,l}X_{k,l+1} +Y_{k,l}Y_{k,l+1})X'_{k,l} Y'_{k,l+1} \]
if $k$ is odd, or
\[ V'_{k,l} := (-1)^{l+1} (X_{k,l}X_{k,l+1} +Y_{k,l}Y_{k,l+1})Y'_{k,l} X'_{k,l+1} \]
if $k$ is even. Each horizontal hopping term $a^\dag_{k,l} a_{k+1,l} + a^\dag_{k+1,l} a_{k,l}$ maps to
\[ H_{k,l} := (X_{k,l}X_{k+1,l} + Y_{k,l}Y_{k+1,l})Z'_{k,l}. \]
The onsite terms remain the same as in the usual JW encoding. Using that $X\otimes X+Y\otimes Y$ can be mapped to $Z \otimes I - I \otimes Z$ by unitary conjugation, time-evolution according to each horizontal term can be implemented with a circuit of 2-qubit gates of depth 4 (which is more efficient than time-evolving according to the $XXZ$ terms and $YYZ$ terms separately). For any term of the form $(X_1 X_2 + Y_1 Y_2)Z_3$, we first map the first 2 qubits to $Z_1 - Z_2$, then perform time-evolution $e^{i \theta Z_1 Z_3}$, $e^{-i \theta Z_2 Z_3}$, then undo the first transformation. The vertical terms are similar, but somewhat more complicated. Now we want to evolve according to a term of the form $(X_1 X_2 + Y_1 Y_2)X_3 Y_4$. We perform the same map on the first 2 qubits; then evolve according to $e^{i \theta Z_1 X_3 Y_4}$; then similarly for $-Z_2$; and then undo the first map. Now the intermediate time-evolution steps each can be implemented using a circuit of depth 3, because they correspond to computing parities of 3 bits each (and some additional 1-qubit gates, which we do not count). However, the parity of qubits 3 and 4 does not need to be recomputed between these time-evolution steps, which saves depth 2; and the unitary operation diagonalising $X_1 X_2 + Y_1 Y_2$ can be performed in parallel with computing this parity. These two optimisations reduce the overall depth complexity of time-evolution according to each vertical term to 4, which is more efficient than implementing the $XXXY$ and $YYXY$ terms separately.

Therefore, the depth required to carry out all time-evolution steps for an arbitrary grid under the Ball-Verstraete-Cirac transformation is $2(4+4) + 1 = 17$, assuming that an arbitrary 2-qubit gate can be implemented in depth 1, and that there are no locality restrictions. This is higher than the cost of executing all time-evolution steps in one layer of the NP ansatz under the Jordan-Wigner transformation for all $n_x \times n_y$ grids such that $\min\{n_x,n_y\} \le 8$. The Ball-Verstraete-Cirac encoding also comes with a significant increase in qubit count (from $2n_x n_y$ to $4(n_x n_y - 1)$~\cite{whitfield16}), as well as an additional cost for preparing the initial state, which we have not considered here.


\subsection{Bravyi-Kitaev superfast encoding}

Bravyi and Kitaev introduced another encoding of fermions as qubits~\cite{bravyi02} which produces $O(1)$-local operators, and which is now known as the Bravyi-Kitaev superfast transformation. In this encoding, one introduces a qubit for every hopping term in $H$ (equivalently, a qubit for each edge in the lattices for each spin), giving an overall system size of $4 n_x n_y - 2 n_x - 2 n_y$ qubits. Then, as described in~\cite{havlicek17}, horizontal hopping terms $a^\dag_j a_k + a^\dag_k a_j$ map to terms of the form
\[ \frac{1}{2} Y_j^{\rightarrow} (Z_j^{\downarrow} Z_k^{\uparrow} - Z_j^{\uparrow} Z_j^{\leftarrow} Z_k^{\rightarrow} Z_k^{\downarrow}), \]
where we follow the notation from~\cite{havlicek17} that arrow superscripts identify qubits in terms of their positions relative to sites $k$ and $j$. Vertical hopping terms $a^\dag_j a_k + a^\dag_k a_j$ map to terms of the form
\[ \frac{1}{2} Y_j^{\uparrow} (Z_k^{\leftarrow} Z_k^{\rightarrow} Z_k^{\uparrow} Z_j^{\leftarrow} Z_j^{\rightarrow} Z_j^{\downarrow} - I). \]
Finally, onsite interactions $n_{k\uparrow} n_{k \downarrow}$ map to terms
\[ \frac{1}{4}(I - Z_k^{\leftarrow} Z_k^{\uparrow} Z_k^{\rightarrow} Z_k^{\downarrow})(I - Z_{k'}^{\leftarrow} Z_{k'}^{\uparrow} Z_{k'}^{\rightarrow} Z_{k'}^{\downarrow}), \]
where $k$ and $k'$ correspond to sites in the spin-$\uparrow$ and spin-$\downarrow$ lattices respectively.

In the horizontal hopping term, all Pauli matrices act on separate qubits with the exception of the $Y_j^{\rightarrow}$ component. Up to local unitary operations on the corresponding qubit, these terms (and the others) can be interpreted as performing rotations conditional on the parities of subsets of bits. The parity of 5 bits that needs to be computed for the $Y_j^{\rightarrow} Z_j^{\uparrow} Z_j^{\leftarrow} Z_k^{\rightarrow} Z_k^{\downarrow}$ part dominates the complexity of the whole evolution, as the part involving 3 qubits ($Y_j^{\rightarrow} Z_j^{\downarrow} Z_k^{\uparrow}$) can be executed in parallel with this. Then the depth required for time-evolution for each hopping term is 6 2-qubit gates (the subroutine comprises a depth-3 circuit of CNOT gates to compute the parity of 5 bits; one 1-qubit rotation gate; and another depth-3 circuit to uncompute). The vertical term is similar, but involves parities of 7 bits, which can also be evaluated in depth 3, giving a depth-6 circuit in total (and noting that the identity term produces a single-qubit gate).

Finally, evolution according to each onsite term can be performed by first storing the parity of the 4 required bits in the lattice for each spin (which requires depth 2), then performing a 2-qubit gate across the two lattices, and uncomputing the first step. The total 2-qubit gate depth is 5.

Note that each of the horizontal and vertical hopping terms across sites $j$ and $k$ involves all qubits adjacent to $j$ and $k$. This implies that (e.g.) considering two horizontal terms across the pairs of sites $(j_1,k_1)$ and $(j_2,k_2)$, if $j_2$ is a neighbour of $k_1$ in a horizontal direction, or $j_2$ is a neighbour of $j_1$ in a vertical direction, there will be qubits that participate in the encoded hopping terms for both of these terms. To avoid these qubits overlapping, all qubits involved in different hopping terms should be distance 2 from each other. This would involve splitting the horizontal (and similarly vertical) hopping terms into 6 groups: by row (even vs.\ odd), and by column (mod 3). A similar issue occurs with the onsite terms, which each involve all qubits neighbouring a particular site. However, here all terms can be implemented using 2 groups.

In total, then, the depth to carry out all time-evolution steps under the Bravyi-Kitaev superfast encoding (under the same assumptions as the previous section) is $2 (6 \times 6) + 2 \times 5 = 82$, which is substantially higher than the Ball-Verstraete-Cirac encoding. We have not attempted to optimise the circuits sketched above, and it is possible that the large overhead from needing to split terms into groups that are implemented separately could be reduced or eliminated, by implementing the required parity operations in a carefully chosen order. If it were possible to implement all groups of commuting horizontal, vertical and onsite terms simultaneously (similarly to the Ball-Verstraete-Cirac encoding) we would achieve a depth of $4 \times 6 + 5 = 29$, which is still worse than the Ball-Verstraete-Cirac encoding.

\section{Implementation on hardware}
\label{sec:implementation}

The description of the EHV ansatz from Section~\ref{sec:var_ansatz} assumes that gates can be implemented across arbitrary pairs of qubits. Most quantum computing architectures have restrictions on their connectivity. These architectures will in general require additional swap operators to move pairs of qubits into positions in which they can interact, and then to move them back again. However, almost all of the gates that are applied in the ansatz take place along the 1D line of the JW ordering; the only other gates are onsite terms. This means that the EHV and NP ans\"atze can be implemented on a $2 \times (n_x n_y)$ nearest-neighbour architecture with no depth overhead per circuit layer. This approach would require an overhead scaling with $n_x$ to measure the vertical hopping terms, and would also require the qubit layout to be particularly ``long and thin'' (or a larger lattice of which this would be a subgraph). In this section we describe alternative approaches to implement the EHV/NP ans\"atze on realistic architectures whose shape is closer to the shape of the grid itself.

\begin{figure*}[t]
    \centering
    \includegraphics[width=0.7\linewidth]{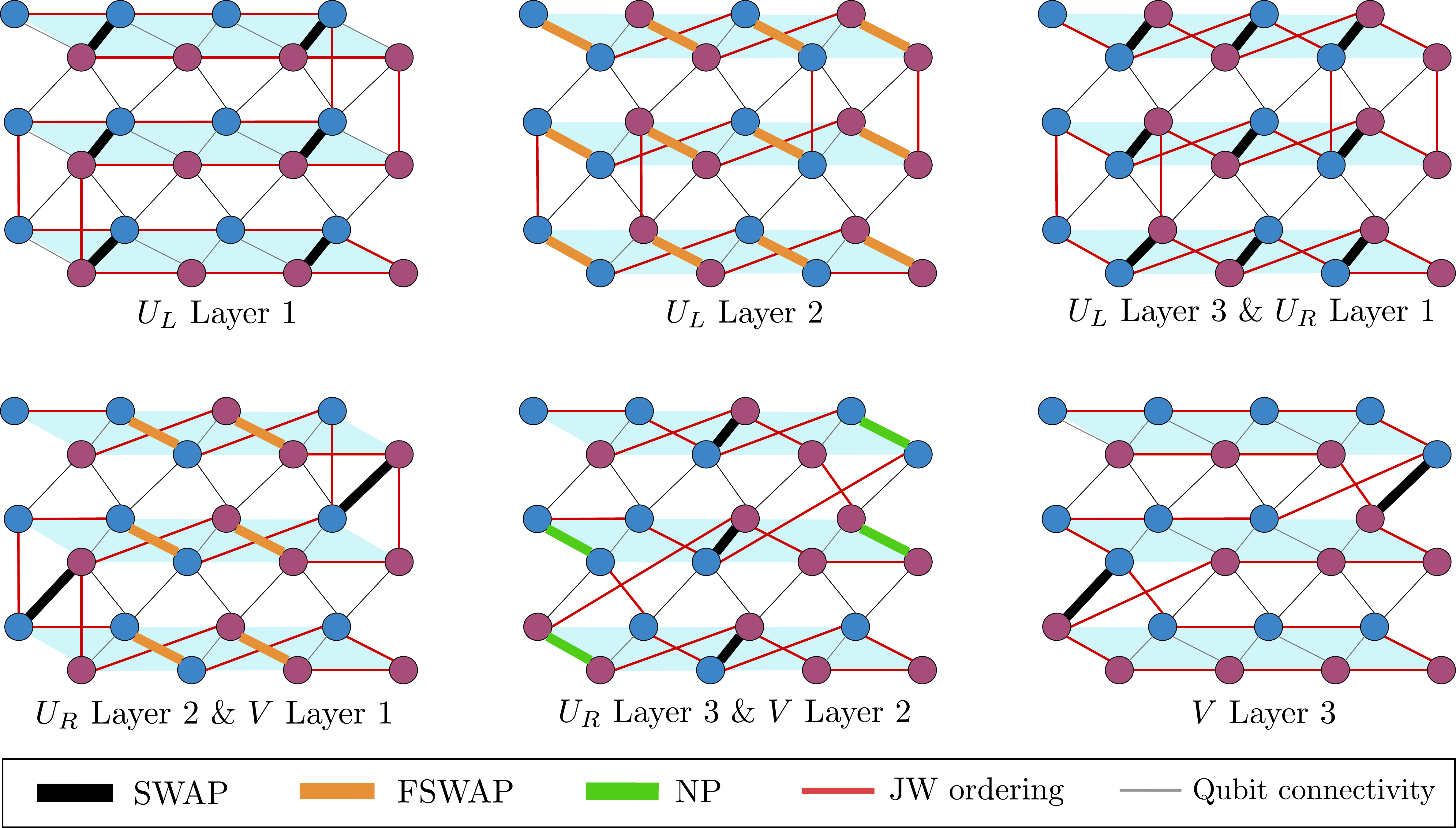}
    \caption{Implementation of the operator $VU_RU_L$ (each split into 3 layers) on the Google Sycamore architecture for even $n_x$, shown here for a $4 \times 3$ grid. The $VU_RU_L$ operator handles only the vertical hopping terms of the NP ansatz, and we remind the reader that the NP ansatz is a generalisation of the HV ansatz. Here we assume that SWAP, FSWAP, and number-preserving ($U_{\text{NP}}$) gates can be implemented in depth 1. Once again the red lines represent the ordering of qubits due to the JW encoding. Observe that during the circuit, the red lines move -- this represents the fact that qubits move physically, whilst retaining the same JW ordering. However, applying an FSWAP gate between two JW-adjacent qubits has the effect of swapping the \emph{ordering} of the two qubits, as well as their physical positions. Hence, FSWAP gates do not alter the relationship between the JW ordering and the physical layout of the qubits, whilst conventional SWAP gates do.}
    \label{fig:Bcone_VRL}
\end{figure*}

Once we have decided on a qubit layout, we can consider the cost of implementing the operator $U_RU_L$ from Section~\ref{sec:var_ansatz}, and how it can be combined with the vertical hopping terms. Since vertical hopping terms are always applied in the same positions (those pairs of qubits that are vertically JW-adjacent), the same operator is used to apply all of them (one round at a time) -- we will call this $V$. The depth of the circuit required to implement one layer of the ansatz will then be determined by the depth of the circuit required to implement $VU_RU_L$, which is repeated $n_x$ times, plus the depth of the circuits used to implement the horizontal hopping and onsite terms. 

\begin{figure}[t]
    \centering
    \includegraphics[width=\linewidth]{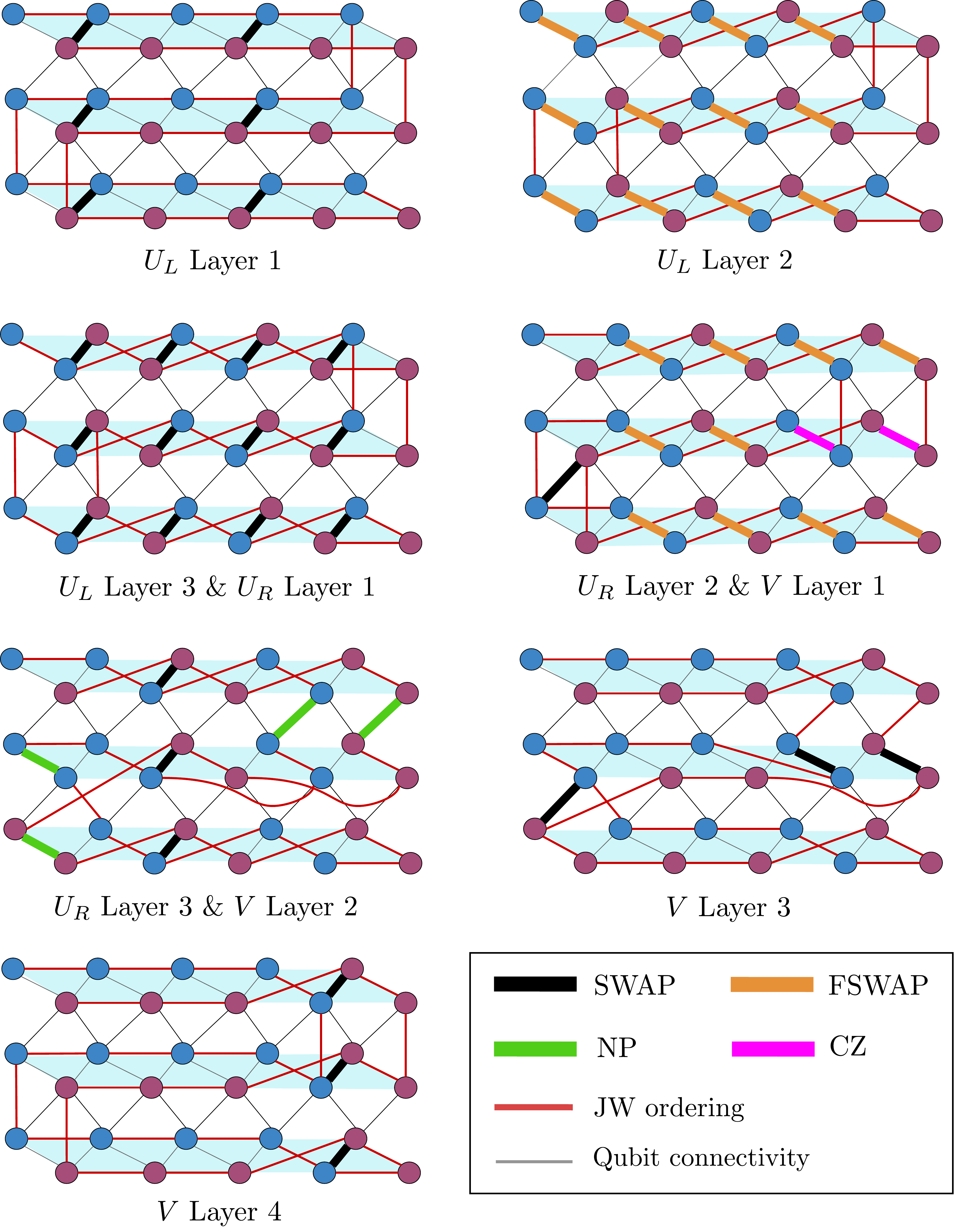}
    \caption{Implementation of the operator $VU_RU_L$ on the Google Sycamore architecture for odd $n_x$, shown here for a $5 \times 3$ grid. Note the $CZ$ gates in the fourth layer of the circuit which is a combination of the FSWAP gate from Layer 2 of $U_R$ and the SWAP gate from Layer 1 of $V$.}
    \label{fig:Bcone_VRL_odd}
\end{figure}

On a nearest neighbour architecture, we could use a qubit layout similar to that described in Figure~\ref{fig:snake_shape}, but where the lattice consists of alternating rows of spin-up and spin-down qubits. In this layout, horizontally JW-adjacent qubits are physically adjacent, but vertically JW-adjacent qubits are not. This means that the operators $U_L$ and $U_R$, which swap horizontally JW-adjacent qubits, can be implemented directly in depth 1 each. The operator $V$ requires that each pair of vertically JW-adjacent qubits are moved so that they become physically adjacent, and then moved back again, which can be achieved using 2 layers of SWAP gates. The first layer of SWAP gates can be implemented in parallel with the $U_R$ operator (for even $n_x$\footnote{For odd $n_x$, some of the SWAP gates can be implemented in parallel with $U_R$, and others with $U_L$. In the end this incurs an extra overhead of only depth 1, using an approach similar to that described in Figure~\ref{fig:Bcone_VRL_odd}.}),  meaning that $VU_RU_L$ can be implemented by a circuit of depth 4 (as was mentioned in Section~\ref{sec:efficient_np}). Also as discussed in Section~\ref{sec:efficient_np}, we can fold the horizontal hopping interactions into the swap network. Finally, all onsite interactions can be implemented in depth 1. This yields a final circuit depth of $4n_x + 1$ per layer.

This approach is quite similar to the swap network used in~\cite{kivlichan18}. There, spin-up and spin-down qubits are adjacent in the JW ordering (with an alternating up, down, down, up, up, \dots\ pattern), as opposed to the alternating rows used here. A depth upper bound of $3\sqrt{2}n_x$ per layer was stated in~\cite{kivlichan18}; it was recently observed by Cai~\cite{cai19} that this can be improved to $4n_x$ using a modified swap network, similar to the one we use here. The depth of $4n_x + 1$ stated here could be decreased to $4n_x$ to match this by combining onsite interactions with SWAP operations, although this would change the ordering of the interactions performed in the ansatz. The alternating approach of~\cite{kivlichan18,cai19} seems to need an additional swap gate at the end when measuring some of the horizontal hopping terms (those corresponding to pairs that are distance 3 in the JW ordering), but it should be possible to remove this by changing the JW ordering for runs that finish by measuring these terms.

The above interlaced approach would result in a physical lattice of shape $n_x \times (2 n_y)$.
However, instead of alternating rows of spin-up and spin-down, we can also place the spin-up lattice physically next to the spin-down lattice.
This results in a lattice of shape $(2 n_x) \times (n_y)$.
The horizontally and vertically JW-adjacent terms are then adjacent on the physical lattice as well, and we can carry out these terms as described in Section~\ref{sec:efficient_np}.
However, the qubits which we want to implement onsite terms across are distance $n_x$ from each other. Using a swap network of depth $n_x - 1$, where the $i$'th layer swaps $i$ pairs of adjacent qubits starting from the middle of each row, we can bring the required qubits next to each other.  We then perform the onsite gate and then use $n_x - 1$ more layers to swap to the original position.
This approach then gives a final depth of $4 n_x - 1$ for even $n_x$, and $4 n_x$ for odd $n_x$ which is a slight improvement on the interlaced approach. Also, the interlaced approach requires an additional layer of swap gates at the end of the algorithm to measure vertical hopping terms, which is not required for the separated approach.

As another example, we consider how to implement the above ansatz efficiently on Google's Sycamore  architecture~\cite{sycamore}. We use the qubit layout described in Section \ref{sec:ferm_encoding}. Once again we are concerned with the depth of the circuit required to implement $VU_RU_L$. In the Sycamore architecture, no JW-adjacent qubits are physically adjacent -- they are all distance 1 away from each other -- and so each of $U_R$, $U_L$, and $V$ must be split into 3 layers each: one to swap qubits into physically adjacent positions; one to carry out the required interaction; and one more to swap the qubits back to their original positions. Many of these layers overlap and can be implemented in parallel. Figure \ref{fig:Bcone_VRL} illustrates how to implement the operator $VU_RU_L$ with a circuit of depth 6 for even values of $n_x$.

Once again, we can fold the horizontal hopping interactions into the swap network, and all onsite interactions can be implemented in depth 1. This yields a final circuit depth of $6n_x+1$ per layer for even values of $n_x$\footnote{Note that there is no dependence on $n_y$. If $n_y < n_x$, we are free to rotate the grid (i.e. by choosing a snake-shaped ordering that travels along the $y$-axis) so that our new grid has $n_y$ columns. Therefore, the circuit depth is more correctly stated as $6\cdot\min\{n_x,n_y\}+1$ for even $n_x, n_y$.}.
For odd values of $n_x$, we lose the ability to implement the vertical hopping terms in parallel with the operator $U_R$, which increases the depth of the final circuit. In figure \ref{fig:Bcone_VRL_odd} we show how to  implement the operator $VU_RU_L$ in depth 7. Here (but not in the even $n_x$ case), the first and last layers can be implemented in parallel, and so we obtain a final circuit depth for the ansatz of $6n_x+2$ per layer, one more than in the even case.

We are now able to compare the effect of different qubit connectivities on circuit depth. These are shown in Table \ref{tab:circuit_depths} in the introduction. An estimate of 2-qubit gate complexity (as opposed to depth) for a complete run of the whole circuit for the efficient version of the HV or NP ansatz follows.

The cost of preparing the initial state is at most $2(N - 1)\lfloor N/2 \rfloor$ gates, where $N = n_x n_y$. Then the cost of the ansatz circuit itself is at most the depth per layer multiplied by the maximal number of 2-qubit gates applied per step of the circuit (which is at most $N$), multiplied by the number of layers. Finally, there is a cost of at most $N$ for the 2-qubit gates required for performing the final measurement.
For example, in the case of a fully-connected architecture, the gate complexity for a circuit with $L$ layers is at most
\be \label{eq:gatecomplexity} (N - 1)N + (2n_x + 1)NL + N \ee
for even $n_x$, and 
\be \label{eq:gatecomplexityodd} 2(N - 1)\lfloor N/2 \rfloor  + (2n_x + 2)NL + N \ee
for odd $n_x$. In the special case of a $2 \times 4$ system with 2 layers, and using a more careful calculation, we obtain a bound of at most 36 gates per layer, giving an upper bound of 136 gates in total. By contrast, the estimate for this case in~\cite{wecker15} was 1000 gates, more than a factor of 7 higher.

\section{The Number Preserving anstaz}
\label{app:np}

In this appendix we will go into details about the choices that can be made when implementing the NP ansatz. As with many ans\"atze, we must specify properties such as starting parameters and initial states.

\subsection{Initial state}
\label{app:init_state}

As well as the ground state of the non-interacting Hubbard model, the NP ansatz also allows a computational basis state with the correct fermionic occupation number as an initial state. All gates in the circuit are fermionic number-preserving, so the VQE method will find the ground state of the Fermi-Hubbard Hamiltonian restricted to the chosen occupation number subspace. This allows a saving in initial complexity compared with starting in the ground state of the non-interacting model (although with an associated penalty in terms of the number of layers required to find the ground state).

The sites we choose to be occupied by fermions can make a significant difference to the complexity at a fixed depth. We ran a number of tests brute forcing all the possible starting states on selected small grid sizes. We found that in many cases the best states reached errors several orders of magnitude better than the worst states, but given the small lattice sizes considered, the pattern for picking these good states remains unclear.

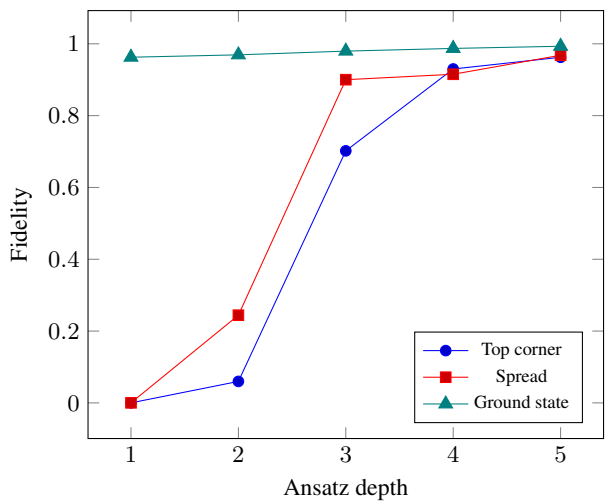
\begin{figure}[t]
    \centering
    \begin{tikzpicture}
    \pgfplotstableread{diagrams/initial_state_graphs/3x3_init_states.txt} \datatable
    \begin{axis}[xlabel=Ansatz depth,ylabel=Fidelity,legend style={font=\scriptsize, at = {(0.96,0.25)}}]
        \addplot table[x=depth, y=top_corner] from \datatable ;
        \addlegendentry{Top corner}
        \addplot table[x=depth, y=spread] from \datatable ;
        \addlegendentry{Spread}
        \addplot[color=teal, mark=triangle*, mark size=3pt] table[x=depth, y=gs] from \datatable;
        \addlegendentry{Ground state}
    \end{axis}
    \end{tikzpicture}
    \caption{Comparison of initial fermion placements against using the ground state as a starting state for a $3 \times 3$ grid occupied by 6 fermions. For the spread out state we occupied both spins for the 3 sites along the main diagonal of the grid. The spread out placement generally performs better than the top corner placement that fills the first 6 orbitals, especially for lower depths. Only starting in the ground state achieves fidelity 0.99, while the others reach around 0.96 in depth 5.}
    \label{fig:init_state}
\end{figure}

An intuitive approach would be to place fermions evenly across the grid, allowing them to quickly spread out.
Then the ground state (if it does indeed correspond to a `spread out' state) can be produced from the initial state using potentially fewer layers of the ansatz circuit. Empirically, we observed that the optimiser performed better with this layout than a na\"ive one where fermions are placed at the top left corner of the grid, although we note that other schemes might yield even better results. Figure \ref{fig:init_state} gives a demonstration for a $3 \times 3$ grid occupied by 6 fermions.

For a $3 \times 3$ grid, ground state of the non-interacting model can be prepared in depth 8 (assuming unrestricted qubit connectivity), whereas each NP ansatz layer requires depth 7. So, in this case, starting with a computational basis state does not seem to be advantageous.
We further remark that the NP ansatz starting from a computational basis state cannot find the true ground state of the non-interacting Hubbard model in the case where the number of fermions with each spin is 1. This is because all computational basis states with Hamming weight 1 are in the null space of this model, and hopping terms preserve this subspace, as we show in Appendix \ref{app:np1fermion}. 

\subsection{Pre-initialising ansatz parameters}

In the main paper, the initial state of the NP ansatz is the non-interacting Hubbard model ground state. However, starting with a computational basis state, the ansatz (and therefore the optimiser) has to do more work to produce something close to the ground state of the full model. 

To reduce the work that the optimiser needs to do, we can first find an ansatz circuit that produces a state close to the ground state of the non-interacting model by classically emulating the VQE procedure. Because we only need to consider a single spin, the number of qubits in the emulation is halved.
For small grid sizes feasible on near-term quantum devices, the non-interacting problem will be tractable on a classical computer. An advantage of classically emulating the procedure (rather than also running these smaller instances on a quantum computer) is that we can use simulated exact measurements.

Once we have performed the optimisation classically, we can pre-initialise the parameters of the full-model ansatz by using the final parameters from the non-interacting model. The intuition is that by allowing the optimisation procedure to begin with a circuit that produces the ground state of the non-interacting model (which we know is a good choice from Figure \ref{fig:repdepths}), it then `only' has to optimise this circuit to produce a ground state of the complete model, having already been pointed in the right direction. 

\begin{figure}[t]
    \centering
    \begin{tikzpicture}
    \pgfplotstableread{diagrams/initial_state_graphs/2x3_pre_init.txt} \datatableone
    \pgfplotstableread{diagrams/initial_state_graphs/3x3_pre_init.txt} \datatabletwo
    \begin{axis}[xlabel=Ansatz depth,ylabel=Fidelity,legend style={font=\scriptsize, at = {(0.96,0.35)}}]
        \addplot table[x=depth, y=normal] from \datatableone ;
        \addlegendentry{$2 \times 3$ Ordinary}
        \addplot table[x=depth, y=pre-init] from \datatableone ;
        \addlegendentry{$2 \times 3$ Pre-initialised}
        \addplot[color = blue, mark=*, mark options=solid, dashed] table[x=depth, y=normal] from \datatabletwo;
        \addlegendentry{$3 \times 3$ Ordinary}
        \addplot[dashed, color = red, mark=square*, mark options=solid] table[x=depth, y=pre-init] from \datatabletwo;
        \addlegendentry{$3 \times 3$ Pre-initialised}
    \end{axis}
    \end{tikzpicture}
    \caption{Comparison of the pre-initialised NP ansatz to the ordinary NP ansatz for $2 \times 3$ occupied by 4 fermions and $3 \times 3$ occupied by 6 fermions. The initial placement of the fermions is spread out (for $2 \times 3$ we fully occupy 2 sites at opposite corners of the grid, $3 \times 3$ is explained in Figure \ref{fig:init_state}). Pre-initialisation improves the results for $2 \times 3$ depth 2, but makes it worse for $3 \times 3$ in all cases. The difference between the ordinary and pre-initialised ansatz reduces as the depth increases; similar behaviour was demonstrated in Figure \ref{fig:init_state}.}
    \label{fig:pre_init}
\end{figure}
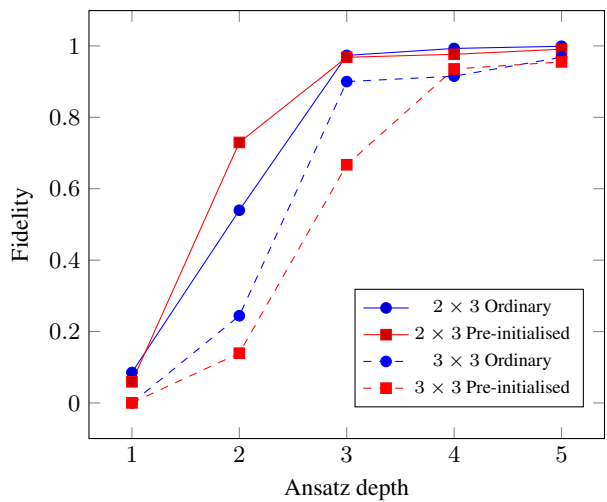

However, it is not clear when this procedure is beneficial as for some grid sizes and depths it causes the ansatz to perform worse. Figure \ref{fig:pre_init} demonstrates this for $2 \times 3$ and $3 \times 3$ grids where the initial placement of the fermions is spread out. We note that different placements change how effective the pre-initialised ansatz is, and that this requires more investigation.

\subsection{Occupation number 1}
\label{app:np1fermion}

Here we show that the NP ansatz starting from a computational basis state cannot find the ground state of the non-interacting Hubbard Hamiltonian, when there is 1 occupied mode.
All computational basis states with Hamming weight 1 are in the null space of the non-interacting Hubbard Hamiltonian. To show that the ground state cannot be found, it is sufficient to prove that time-evolution according to hopping terms preserves this subspace.

In a system with $N$ modes, any state which is a linear combination of occupation number 1 basis states can be written as $\sum_{k=1}^N \alpha_k \ket{e_k}$ for some coefficients $\alpha_k$, where $e_k$ is the vector with Hamming weight 1 whose $k$'th entry is 1.
Within this $N$-dimensional space, the hopping term $(X_iX_j + Y_i Y_j)/2$ (where $i$ and $j$ are adjacent in the Jordan-Wigner ordering) acts as an X gate within the 2-dimensional subspace $\operatorname{span}\{\ket{e_i},\ket{e_j}\}$. Write $X_{ij}$ for this gate.
A state with Hamming weight 1 is contained within the null space of the hopping term between modes $i$ and $j$ (assuming that $i$ and $j$ are adjacent in the Jordan-Wigner ordering) if
\begin{eqnarray*}
0 &=& \left(\sum_{k=1}^N \alpha_k^* \bra{e_k}\right) X_{ij} \left(\sum_{l=1}^N \alpha_l \ket{e_l} \right)\\
&=& \alpha_i^* \alpha_j + \alpha_j^* \alpha_i\\
&=& 2\operatorname{Re}(\alpha_i^* \alpha_j).
\end{eqnarray*}

Consider an arbitrary 3-dimensional subspace corresponding to adjacent modes $i$, $j$, $k$ in the Jordan-Wigner ordering. Then
\begin{multline*} e^{i\theta X_{ij}}(\alpha \ket{e_i} + \beta \ket{e_j} + \gamma \ket{e_k})\\
= (\alpha \cos \theta + i \beta \sin \theta) \ket{e_i} + (i \alpha \sin \theta + \beta \cos \theta) \ket{e_j} + \gamma \ket{e_k}\\
=: \alpha' \ket{e_i} + \beta' \ket{e_j} + \gamma' \ket{e_k}.
\end{multline*}
To show that this state is contained within the null space of all hopping terms, it is sufficient to show that $\operatorname{Re}((\alpha')^*\beta') = \operatorname{Re}((\gamma')^*\beta') = 0$.

The former claim is immediate as the initial state is in the null space of $X_{ij}$. For the latter claim, we have
\begin{eqnarray*}
\operatorname{Re}((\gamma')^* \beta') &=& \operatorname{Re}(\gamma^*(i \alpha\sin\theta + \beta \cos \theta ) )\\
&=& \cos \theta \operatorname{Re}(\gamma^* \beta) - \sin \theta \operatorname{Im}(\gamma^* \alpha).
\end{eqnarray*}
We have $\operatorname{Re}(\gamma^* \beta) = 0$ as the initial state is in the null space of $X_{jk}$. To see that $\operatorname{Im}(\gamma^* \alpha) = 0$, write $\alpha = r_\alpha e^{i s_\alpha}$, and similarly for $\beta$, $\gamma$. Then, as $\alpha^* \beta$ and $\beta^* \gamma$ are imaginary from the same null space constraint, we have that $s_\beta - s_\alpha$ and $s_\gamma - s_\beta$ are in the set $\{\pm \pi/2, \pm 3\pi/2\}$. So $s_\gamma - s_\alpha$ must be an integer multiple of $\pi$, implying that $\gamma^* \alpha$ is real.


\section{Simulation choices}

This appendix summarises the reasoning behind some choices that were made in our tests, and presents additional results for other regimes.

\subsection{Effect of choice of \texorpdfstring{$U$}{U} parameter}
\label{sec:U_param}
Throughout this work, we fixed the weight $U$ of the onsite term in the Hubbard Hamiltonian (\ref{eq:hubbard}) to 2, as was also done in~\cite{wecker15}. To justify this, we considered three grid sizes ($2\times 2$, $1\times 6$ and $3\times 3$) and evaluated the fidelity achieved for different choices of $U$ by optimising using L-BFGS with exact energy measurements within the EHV ansatz, at the same depth for which the $U=2$ case achieves fidelity $>0.99$. This gives a measure of the difficulty of finding the ground state. The results are shown in Figure \ref{fig:urange}. One can see that the fidelity decreases as $U$ increases, as expected given that the ansatz begins in the ground state of the $U=0$ model. However, the final fidelity achieved continues to be quite high for all $U \le 4$.

\begin{figure}[t]
    \centering
    \begin{tikzpicture}
    \pgfplotstableread{diagrams/results/urange_depth_to_99.txt} \datatable
    \begin{axis}[xlabel=U,ylabel=Depth to 0.99 fidelity, legend style = {font=\scriptsize, at = {(0.3,0.95)}},  yminorgrids=true, ymajorgrids=true, minor y tick num=4, major grid style={thick}]
        \addplot[color=blue, mark=no mark, thick] table[x=U, y=2x2] from \datatable ;
        \addlegendentry{$2 \times 2$}
        \addplot[color = red, mark=no mark, thick] table[x=U, y=1x6] from \datatable ;
        \addlegendentry{$1 \times 6$}
        \addplot[color=teal, mark=no mark, thick] table[x=U, y=3x3] from \datatable;
        \addlegendentry{$3 \times 3$}
    \end{axis}
    \end{tikzpicture}
    \caption{Depth of the Efficient Hamiltonian Variational ansatz required to reach 0.99 fidelity with the ground state of the Hubbard model as a function of $U$ with grid sizes $2 \times 2$, $1 \times 6$ and $3 \times 3$.}
    \label{fig:urange_depth_to_99}
\end{figure}
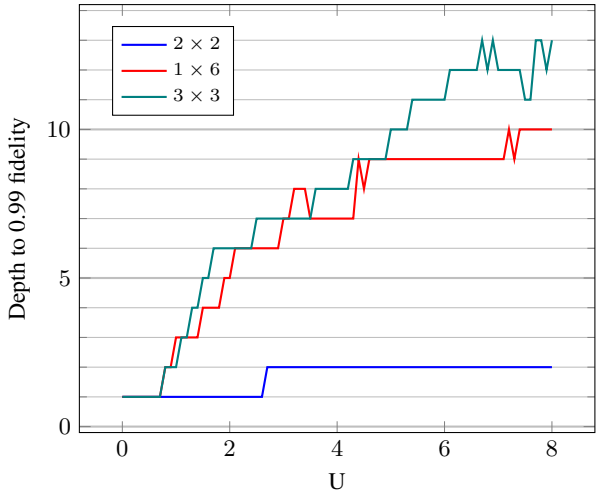

Figure \ref{fig:urange_depth_to_99} demonstrates the minimal depth of the EHV ansatz required to reach 0.99 fidelity as $U$ varies. In general the depth required increases as $U$ does, which is to be expected as we begin in the $U=0$ ground state. As we can see from Figure \ref{fig:urange_depth_to_99}, to get to the physically interesting intermediate coupling regime $U = 4$, where classical methods experience significant uncertainties~\cite[Table V]{leblanc15}, only requires 1 or 2 extra ansatz layers from $U = 2$. However, the more strongly correlated model $U = 8$ requires roughly double the ansatz layers. 

We remark that, when optimising with realistic measurements, the statistical uncertainty in the energy measurements is likely to increase linearly with $U$. This is because the energy is measured by summing several measurement results, some of which are scaled by a $U$ factor. Thus going from $U=2$ to $U=8$ (for example) would likely require 16 times more measurements to achieve the same level of statistical uncertainty.

\begin{figure}[t]
    \centering
    \begin{tikzpicture}
    \pgfplotstableread{diagrams/results/urange.txt} \datatable
    \begin{axis}[xlabel=U,ylabel=Fidelity,legend style={font=\scriptsize, at = {(0.96,0.95)}}, yticklabel style={/pgf/number format/.cd,fixed zerofill,precision=3}]
        \addplot[color=blue, mark=no mark, thick] table[x=U, y=2x2] from \datatable ;
        \addlegendentry{$2 \times 2$}
        \addplot[color = red, mark=no mark, dashed, thick] table[x=U, y=1x6] from \datatable ;
        \addlegendentry{$1 \times 6$}
        \addplot[color=teal, mark=no mark, dotted, very thick] table[x=U, y=3x3] from \datatable;
        \addlegendentry{$3 \times 3$}
    \end{axis}
    \end{tikzpicture}
    \caption{The final fidelity achieved with varying $U$ for a $2\times 2$ grid (at depth 1), $1\times 6$ (at depth 5), and $3 \times 3$ (at depth 6), using the EHV ansatz, simulated exact energy measurements, and the L-BFGS optimiser. $U$ incremented in steps of size 0.1.}
    \label{fig:urange}
\end{figure}
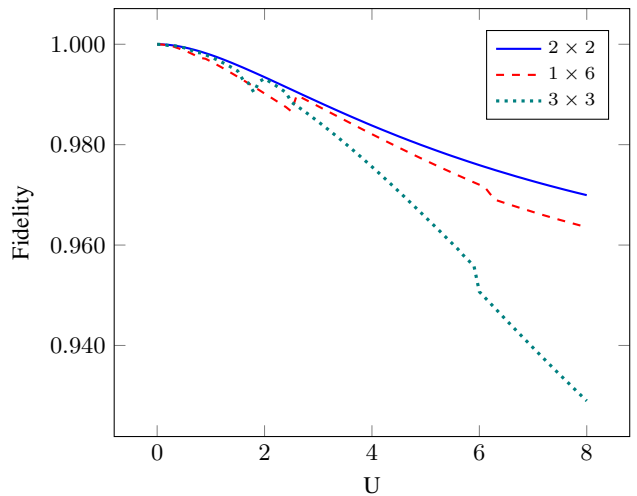

\subsection{The half-filled regime}

While we were mostly concerned with finding the ground state of the original Hamiltonian presented in (\ref{eq:hubbard}), solutions of certain restricted cases can be of interest as well. A prominent restriction is that of ``half-filling'', where the number of fermions in the lattice is exactly half of the number of sites. This case is easier to solve classically due to the lack of a sign problem~\cite{leblanc15}, enabling quantum Monte Carlo methods to succeed. However, the special physical and mathematical characteristics of the half-filled regime make it an important one in which to also benchmark VQE methods.

\begin{figure}[t]
    \centering
    \begin{tikzpicture}
    \pgfplotstableread{diagrams/results/half_fill_depthscaling.txt} \datatable
    \begin{semilogyaxis}[xlabel=Ansatz depth,legend style={font=\scriptsize, at = {(0.96,0.95)}}, ymode=log, grid=both]
        \addplot table[x=depth, y=infid] from \datatable ;
        \addlegendentry{1 - Fidelity}
        \addplot table[x=depth, y=energy] from \datatable ;
        \addlegendentry{Energy error}
        \addplot[color=teal, mark=triangle*, mark size=3pt] table[x=depth, y=double_occ] from \datatable;
        \addlegendentry{Double occupancy error}
    \end{semilogyaxis}
    \end{tikzpicture}
    
    \caption{Infidelity, absolute error with the actual ground state and absolute error with the true double occupancy for various depths of the Efficient Hamiltonian Variational ansatz for the $2 \times 4$ half-filled Hubbard model.}
    \label{fig:half_fill_scaling}
\end{figure}
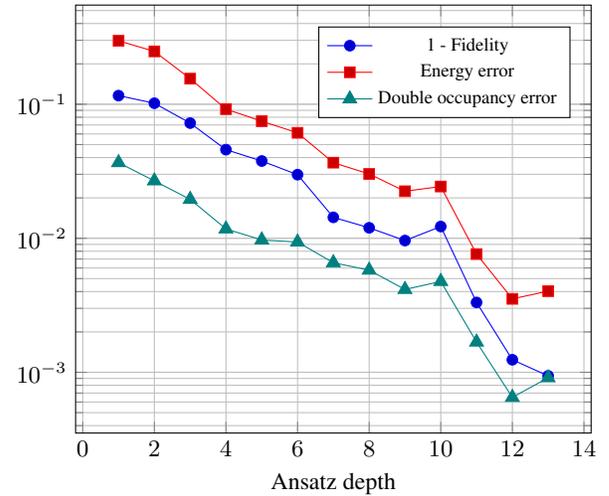

The performance of our algorithm in terms of depth to high-fidelity solution can be seen in Fig.~\ref{fig:half_fill_depths}; we can see that the depths required in the half-filling case are comparable to depths required to find the ground state of the full Hamiltonian for the same ansatz.
In Fig.~\ref{fig:half_fill_scaling}, we can see how the infidelity, absolute error with the actual ground state, and absolute error with the true double occupancy of the ground state changed with depth for an example grid size of 2x4, also at half-filling.
While the optimisation is carried out by minimising the energy, we can see that the infidelity and the error in the double occupancy follow a very similar trend to the error in the ground energy.
This gives us reason to believe that energy is a good figure-of-merit to optimise on, even if a different property of the ground state (such as double occupancy) might be the one we are interested in. The situation is similar away from half-filling.

Another peculiarity about the half-filled case is that degeneracy in the ground states of the non-interacting Hamiltonian, which is the initial state for the EHV ansatz, is common.
If the degeneracy is low enough (only a few states), then trying out each of the degenerate states as the initial state might be feasible.
However, in some of the lattices with higher degeneracy we tried a few different solutions to arrive at a successful initial state.
For the results presented in Fig.~\ref{fig:half_fill_depths}, the initial states were generated as follows: if there was no degeneracy then the choice was the single ground state; for grid size 2x2, one of the hopping terms in the Hamiltonian (between sites top left two sites) was altered by $\epsilon = 0.0001$ allowing for a splitting between the degenerate states; for grid sizes 2x5, 3x3, a superposition over all the degenerate states was created and the weights of each of the states were added as parameters to be optimized over in the main optimization; for all other degenerate states, a manual selection of initial state was carried out by trial-and-error.

\begin{figure}[t]
    \centering
    \begin{tikzpicture}
    \begin{axis}[xlabel=Grid height $n$,ylabel=Depth to 0.99 fidelity, legend style = {font=\scriptsize, at = {(0.5,1.03)}, anchor=south, legend columns={3}, /tikz/every even column/.append style={column sep=0.35cm}},  yminorgrids=true, ymajorgrids=true, minor y tick num=4, major grid style={thick}]
    
    \addplot[color=red, mark=square*, mark size=1pt, thick] coordinates {(2, 1) (3, 3) (4, 3) (5, 5) (6, 4) (7, 7) (8, 6) (9, 9) (10, 8) (11, 11) (12, 10)};
    \addlegendentry{$1\times n$ half-fill};
    
    \addplot[color=teal, mark=square*, mark size=1pt, thick] coordinates {(2, 5) (3, 5) (4, 9) (5, 12) (6, 17)};
    \addlegendentry{$2\times n$ half-fill};
    
    \addplot[color=blue, mark=square*, mark size=1pt, thick] coordinates {(3, 7)};
    \addlegendentry{$3\times n$ half-fill};

    \addplot[color=red, mark=x, thick, dotted, mark options=solid] coordinates {(2, 1) (3, 3) (4, 3) (5, 5) (6, 5) (7, 7) (8, 7) (9, 7) (10, 9) (11, 10) (12, 9)};
    \addlegendentry{$1\times n$ overall};
    \addplot[color=teal, mark=x, thick, dotted, mark options=solid] coordinates {(2, 1) (3, 3) (4, 11) (5,11) (6,18)};
    \addlegendentry{$2\times n$ overall};
    \addplot[color=blue, mark=x, thick, dotted, mark options=solid] coordinates {(3, 6)};
    \addlegendentry{$3\times n$ overall};

    \end{axis}
    \end{tikzpicture}
\caption{Depth of the Efficient Hamiltonian Variational ansatz required to reach 0.99 fidelity with the ground state of the half-filled Hubbard model $(t=1, U=2)$. Comparison with depth required to find the overall ground state (data reproduced from Figure \ref{fig:repdepths}). }
\label{fig:half_fill_depths}
\end{figure}
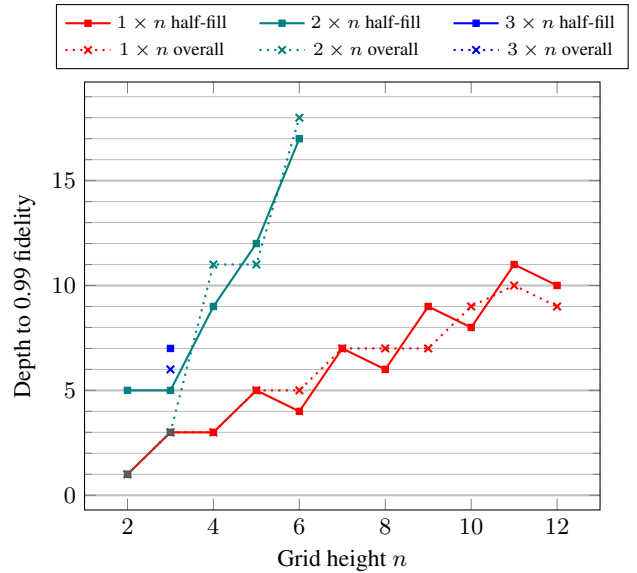

\subsection{Characterising statistical noise in the ansatz circuits}

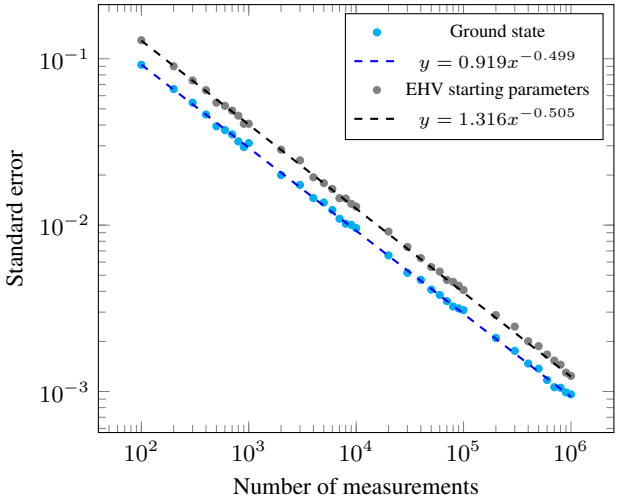
\begin{figure}[t]
    \centering
    \begin{tikzpicture}
    \pgfplotstableread{diagrams/results/stderror.txt} \datatable
    \begin{loglogaxis}[xlabel=Number of measurements,ylabel=Standard error, clip mode=individual, legend style = {font=\scriptsize}]
        \addplot[only marks, cyan, mark size = {0.15em}] table[x=nmeas, y=gs] from \datatable ;
        \addlegendentry{Ground state};
        \addplot[no marks, dashed, blue, thick] coordinates {(100, 9.219e-02) (1e6, 9.284e-04)};
        \addlegendentry{$y=0.919x^{-0.499}$}
        \addplot[only marks, gray, mark size = {0.15em}] table[x=nmeas, y=hv] from \datatable ;
        \addlegendentry{EHV starting parameters};
        \addplot[no marks, dashed, black, thick] coordinates {(100, 1.286e-01) (1e6, 1.229e-03)};
        \addlegendentry{$y=1.316x^{-0.505}$};
    \end{loglogaxis}
    \end{tikzpicture}
    \caption{Statistical error in approximating the energy with respect to the number of measurements made. Results are shown for a $2 \times 2$ grid and the starting parameters chosen for EHV are $1/d$ where $d$ is the depth of the ansatz (6 in this case). Each point on the graph is the standard deviation of 1,000 samples, where each sample is the error in the estimated energy achieved using $m$ measurements.}
    \label{fig:statistical_error}
\end{figure}

In Figure \ref{fig:statistical_error}, we present numerical results that justify performing $10^4$ measurements on each term in the Hamiltonian to estimate the energy to an accuracy of $\sim 10^{-2}$. 
The statistical error on the state is larger when the circuit which produces it is not generating the ground-state (i.e.\ an eigenstate) of the Hamiltonian. Note that the lines of best fit in the figure show that the standard error goes down like $1/\sqrt{m}$, where $m$ is the number of measurements, as is expected.

\section{Preparing the initial state of the non-interacting Hubbard Hamiltonian}
\label{app:fft}

This appendix compares the complexities of different methods for preparing the ground state of the non-interacting Hubbard Hamiltonian ((\ref{eq:hubbard}) with $U=0$): first, approaches to implement the 2D fermionic Fourier transform (FFT) on rectangular grids of size $n_x \times n_y$; and then an approach based on Givens rotations~\cite{jiang2018quantum}. We will see that the latter is the most efficient for small grid sizes, while for large grid sizes an FFT algorithm of~\cite{jiang2018quantum} is superior. The most efficient implementation of the full FFT for small grid sizes is the approach based on FSWAP networks.


\subsection{Na\"ive approach to implementing the FFT}\label{sec:FFT_naiive}
The na\"ive approach to implementing the FFT first separates it into horizontal and vertical components, $\mathcal{F}_x$ and $\mathcal{F}_y$. The terms $\mathcal{F}_x$ and $\mathcal{F}_y$ are products of commuting terms that involve qubits only in the same row and column, respectively. To implement $\mathcal{F}_x$, we apply the 1D Fourier transform on all rows in parallel. To implement $\mathcal{F}_y$, it is necessary to implement the 1D Fourier transform on all columns, but with the appropriate parity corrections ($Z$-strings) attached to each Givens rotation performed. The parity corrections prevent us from implementing this part of the circuit on all columns in parallel, since the corrections span across multiple columns. Assuming that we don't implement any of the $Z$-strings acting on the same row in parallel, and we use a simple 1D nearest-neighbour circuit for computing the parity corrections, then the depth of any FFT circuit implemented na\"ively in this way will be 
\[
    T_{\mathcal{F}}(n_x) + T_{\mathcal{F}}(n_y)\cdot \sum_{i=1}^{n_x} 2(n_x-i)+1 =  T_{\mathcal{F}}(n_x) + T_{\mathcal{F}}(n_y)\cdot n_x^2,
\]
where $T_{\mathcal{F}}(n)$ is the depth of the circuit implementing the 1D fermionic Fourier transform on $n$ qubits. For an $n\times n$ lattice, the depth is thus $\Theta(n^3)$, assuming the depth of the 1D FFT is $O(n)$\footnote{In a fully-connected architecture, parallel circuits could be used to implement the parity corrections; however, these would still not be competitive with the best complexity we find below using swap networks.}.

\subsection{Asymptotically efficient implementation of the FFT}
\label{sec:FFT_asym}
Jiang et al.~\cite{jiang2018quantum} described a method to implement the FFT on a 2D qubit array of size $n_x \times n_y =: N$ with $O(\sqrt{N})$ depth and $O(N^{3/2})$ gates. As in the previous section, the general approach is to factor the FFT into its horizontal and vertical components $\mathcal{F} = \mathcal{F}_x\mathcal{F}_y$. 

Under the Jordan-Wigner transform, the horizontal part is straightforward to implement. Indeed, we can implement the 1D Fourier transform in parallel for all rows, without the need for parity corrections. However, the vertical component is much harder to implement because of the non-local parity operators required to correctly implement 2-qubit interactions between neighbouring qubits in a column. The approach developed in \cite{jiang2018quantum} is to decompose the vertical term as 
\[
	\mathcal{F}_y = \Gamma^\dag \mathcal{F}_y^b \Gamma,
\]
where $\mathcal{F}_y^b$ is the vertical component without the parity operators (i.e. the 1D Fourier transform), and $\Gamma = \Gamma^\dag$ is a diagonal (in the computational basis) unitary that `re-attaches' the parity operators. $\mathcal{F}_y^b$ can be implemented using the same circuit as for $\mathcal{F}_x$, but applied to all columns in parallel. The circuit for $\Gamma$ is more complicated.

The operator $\Gamma$ can be implemented by attaching an additional qubit per row, and then using these to keep track of the parities of the qubits in their corresponding row. The general approach is roughly as follows (for details, see~\cite{jiang2018quantum}):
\begin{enumerate}
\item Convert each column to the parity basis via a sequence of CNOT gates. 
\item Move the ancilla qubits to the left whilst updating their parity, using a SWAP gate, followed by a CNOT between the ancilla and the `system' qubit now to its right. As the qubits move to the left, we apply a sequence of CZ gates to update the phases of the system qubits. 
\item Once all qubits reach the left-hand side, undo the conversion to the parity basis by undoing the CNOT gates.
\item Move the ancilla qubits rightwards by applying the CNOT and SWAP gates in reverse. At each step apply some more CZ gates to update the parities of the system qubits correctly. 
\end{enumerate}

Every step requires a circuit of depth $O(\sqrt{N})$ with $O(N)$ gates. Here we will attempt to calculate the constants associated with this asymptotic scaling.

\begin{figure}[t]
    \centering
    \includegraphics{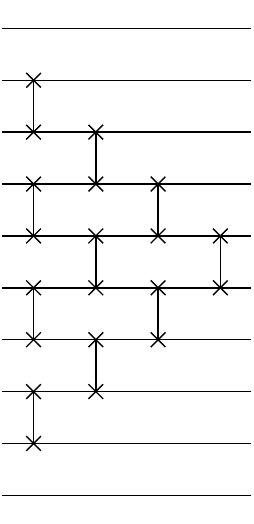}
    \caption{Circuit to bring all odd-numbered rows and all even-numbered rows together. }
    \label{fig:swaps}
\end{figure}

\begin{enumerate}
\item Step 1 is a transformation to the parity basis. This circuit requires $n_y-1$ gates per column, and therefore $n_x(n_y-1) = N-n_x$ gates in total, with a depth of $n_y-1$ (see Figure \ref{fig:parity_conversion})\footnote{This depth could be reduced to $O(\sqrt{n_y})$ on a nearest-neighbour architecture, or $O(\log n_y)$ on a fully-connected architecture (Craig Gidney, personal communication). This would reduce the grid size slightly at which this approach starts to outperform its competitors.}.

\item The circuit in step 2 is more complicated to implement efficiently. To implement these gates as they are described in~\cite{jiang2018quantum} on a nearest neighbour architecture, we would need to perform a number of swap operations to bring the system and ancilla qubits together, requiring two swap operations per gate. Luckily, we can avoid paying double for these gates by re-ordering the SWAP and CNOT gates used to move the ancilla qubits to the left. 

\begin{figure}[t]
    \centering
    \includegraphics{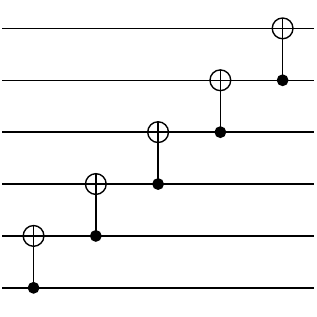}
    \caption{Circuit to transform qubits in a column to the parity basis.}
    \label{fig:parity_conversion}
\end{figure}

By using this re-ordering approach, the total number of gates required to implement the step is $\frac{3n_y}{2}\cdot n_x = \frac{3N}{2}$, and the total depth is $4n_x$.

This calculation ignores the fact that there are vertical CZ gates acting on non-adjacent rows. Here we have two options. One option is to move the qubits in all odd-numbered rows so that they are all adjacent to each other before step 2, and then move them back afterwards. Using the circuit in Figure \ref{fig:swaps}, this adds an additional overhead of $n_x\cdot\left(\frac{n_y}{2}-2\right)\left(\frac{n_y}{2}-1\right)$ gates with a depth of $n_y-2$ (for both doing and undoing the circuit). 

Alternatively, we could just apply SWAP gates as and when we need them. This would involve applying two SWAP gates per vertical CZ operation in this step. We can't apply any of them in parallel with any of the CZ operations, and so we have a total overhead of $n_x\cdot n_y$ gates and an increased depth of $n_x$. Hence, we always save a constant depth of 2 by using the first approach and for $4 \leq n_y \leq 13$, it uses fewer gates. 

Choosing the first approach for implementing CZ gates on non-adjacent rows, step 2 can be implemented using a circuit with depth $4n_x+n_y-2$. If we are allowed to apply gates to arbitrary pairs of qubits, then the circuit depth reduces to just $4n_x$.

\item Step 3 uses the same gates as step 1, except in reverse and with $n_y-1$ additional CNOT gates for the `ancilla column', and therefore requires a circuit of $(n_x+1)(n_y-1)$ gates with a depth of $n_y-1$. 

\item Step 4 is similar to step 2, but somewhat simpler. We move the ancilla qubits to the right using CNOT and SWAP gates, whilst applying local CZ gates at each time step. The number of gates required to move the qubits to the right is $2n_xn_y$, and the number of CZ gates that need to be applied is $n_xn_y$, giving a total of $3n_xn_y$. We can apply all gates in parallel for each row, giving a total depth of $2n_x+2n_x=4n_x$.

Once again, using a nearest neighbour architecture and the first approach mentioned in Step 2 would increase this circuit depth to $4n_x + n_y - 2$.
\end{enumerate}

Putting all these steps together gives us a total circuit depth to implement $\Gamma$ of 
\[
n_y-1 + n_y-1 + 4n_x + 4n_x = 2n_y + 8n_x - 2
\]
if we are allowed to apply gates to arbitrary pairs of qubits, and a depth of 
\[
n_y-1 + n_y-1 + 4n_x+n_y-2 + 4n_x+n_y-2 = 4n_y + 8n_x - 6
\]
if we use a nearest neighbour architecture. 
Suppose that the depth of the circuit used to implement the 1D FFT on $n$ qubits is $T_{\mathcal{F}}(n)$. Then combining the three stages described above: $\mathcal{F}_x$, $\mathcal{F}^b_y$, and $\Gamma$ and $\Gamma^\dag$, we obtain a final circuit depth of 
\[
    T_{\mathcal{F}}(n_x) + T_{\mathcal{F}}(n_y) + 2(8n_x + 2n_y - 2).
\]
If we are restricted to an architecture that only allows interactions between neighbouring qubits, the depth of the circuit required to implement the FFT is 
\[
    T_{\mathcal{F}}(n_x) + T_{\mathcal{F}}(n_y) + 2(8n_x + 4n_y - 6).
\]
However, when we combine all stages of the FFT circuit together, there are some overlaps that are not accounted for in the above analysis. This means that the actual (optimised) circuit depth will be slightly less than predicted. In~\cite{jiang2018quantum}, the authors show that the 1D Fourier transforms can be implemented with circuits of depth $T_{\mathcal{F}}(n_x) = n_x-1$ and $T_{\mathcal{F}}(n_y) = n_y-1$.
The table \ref{tab:fft_comparison} shows the actual vs. predicted depths of the FFT circuit for a number of (square) grid sizes on an unrestricted architecture. From these numbers it appears that parallelisation of the stages gives us a depth saving of $3(n/2-2)$ for an $n \times n$ grid.

\begin{table*}[t]
\centering
\begin{tabular}{|c|c|c|c|c|c|}
\hline
Grid size & \begin{tabular}[c]{@{}c@{}}Asym. efficient \\ (predicted)\end{tabular} & \begin{tabular}[c]{@{}c@{}}Asym. efficient\\ (actual)\end{tabular} & Swap network & \begin{tabular}[c]{@{}c@{}}Swap network\\ (modified)\end{tabular} & Givens rotations \\ \hline
$4 \times 4$ & 82 & 82 & 54 & 27 & 15 \\
$6 \times 6$ & 126 & 123 & 112 & 65 & 35 \\
$8 \times 8$ & 170 & 164 & 265 & 119 & 63 \\
$10 \times 10$ & 214 & 205 & 383 & 189 & 99 \\
$12 \times 12$ & 258 & 246 & 636 & 275 & 143 \\
$14 \times 14$ & 302 & 287 & 814 & 377 & 195 \\
$16 \times 16$ & 346 & 328 & 1167 & 495 & 255 \\
$18 \times 18$ & 390 & 369 & 1492 & 629 & 323 \\ \hline
\end{tabular}
\caption{Comparison of FFT circuit depths and directly preparing the Slater determinant using Givens rotations for a variety of $n \times n$ grids.}
\label{tab:fft_comparison}
\end{table*}

\subsection{Fermionic swap networks for the FFT}\label{sec:swap_networks}
To avoid needing to implement the phase corrections from the previous section, we could instead use the notion of a FSWAP network~\cite{kivlichan18}. Here, we use FSWAP operators to move qubits next to each other so that they can interact without the need for parity corrections. Crucially, these swap operators correctly maintain the relative phases between qubits required by the JW ordering. The idea is to apply a number of layers of 2-qubit FSWAP gates so that by the time we are done every qubit has been adjacent to every other qubit, enabling it to interact without needing to worry about phase corrections due to the JW encoding. This notion can be extended to a 2D grid of spin orbitals. Following~\cite{kivlichan18}, by using a total of $3\sqrt{N/2}$ FSWAP operators we can implement all vertical and horizontal gates in the FFT using gates that can be implemented by nearest neighbour interactions.
These swap operators remove the need to implement the $Z$-strings required to correctly simulate the vertical hopping terms under the JW transform. 

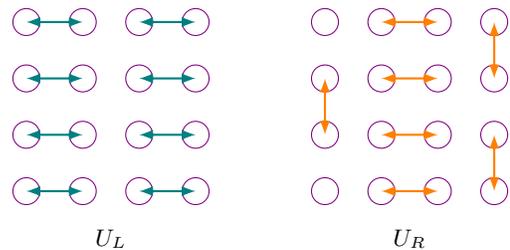
\begin{figure}[t]
    \centering
    \begin{minipage}[t]{0.45\linewidth}
    \begin{tikzpicture}[scale = 0.75, node distance = 0.2cm]
        \foreach \x in {0,...,3}
        \foreach \y in {0,...,3}
        {
            \node[qubit2] (\x-\y) at (\x,\y){};
        }
        \draw[-, draw=white] (1-0.south) to [out=290,in=250] node [below] {$U_L$} (2-0.south);
        
        \foreach \mynode/\joinednode in {0-0/1-0, 0-1/1-1, 0-2/1-2, 0-3/1-3, 2-0/3-0, 2-1/3-1, 2-2/3-2, 2-3/3-3}
        {
            \draw[<->, >=latex, thick, draw=teal] (\mynode.center) -- (\joinednode.center);
        }
    \end{tikzpicture}
    \end{minipage}
    \begin{minipage}[b]{0.45\linewidth}
    \begin{tikzpicture}[scale = 0.75]
        \foreach \x in {0,...,3}
        \foreach \y in {0,...,3}
        {
            \node[qubit2] (\x-\y) at (\x,\y){};
        }
        \draw[-, draw=white] (1-0.south) to [out=290,in=250] node [below] {$U_R$} (2-0.south);

        \foreach \mynode/\joinednode in {1-0/2-0, 1-1/2-1, 1-2/2-2, 1-3/2-3, 3-3/3-2, 0-1/0-2, 3-0/3-1}
        {
            \draw[<->, >=latex, thick, draw=orange] (\mynode.center) -- (\joinednode.center);
        }
    \end{tikzpicture}
    \end{minipage}
    \caption{Action of sets of fermionic swaps $U_L$ and $U_R$ on a $4 \times 4$ grid of qubits, using the swap network of~\cite{kivlichan18}. The ordering of the qubits is the snake ordering in the main paper.}
    \label{fig:fft_UR_UL}
\end{figure}

The approach of~\cite{kivlichan18} is based on a repeated pattern of fermionic swaps denoted $U_L$ and $U_R$, where (unlike the definition in the main text of the present work) these occur along the ``snake'' ordering in the JW encoding (see Figure \ref{fig:fft_UR_UL}). Using these, one is able to bring spin-orbitals from adjacent rows next to each other in the canonical ordering so that the hopping term may be applied locally. First, one applies $U_L$. This will enable application of the first vertical hopping term that could not previously be reached. Then, one should repeatedly apply $U_LU_R$. After each application of $U_LU_R$, new vertical hopping terms become available until one has applied $U_LU_R$ a total of $\sqrt{N/8}-1$ times. At that point, one needs to reverse the series of swaps until the orbitals are back to their original locations in the canonical ordering. After this, applying $U_LU_R$ will cause the qubits to circulate in the other direction. This should be repeated a total of $\sqrt{N/8}-1$ times to make sure that all neighbouring orbitals are adjacent at least once. The total number of layers of fermionic swaps required for the whole procedure is $3\sqrt{N/2}$.

To see how this swap network works for the FFT, we need to consider the structure of the 1D FFT circuit. If we use the approach from Jiang et al.~\cite{jiang2018quantum} to implement the 1D FFT, then in the example of the $4\times 4$ grid, there are two stages to the circuit: first we apply Givens rotations between all vertically adjacent qubits in the 2\textsuperscript{nd} and 3\textsuperscript{rd} rows, and then we apply Givens rotations between all vertically adjacent qubits in the 1\textsuperscript{st} and 2\textsuperscript{nd}, and 3\textsuperscript{rd} and 4\textsuperscript{th} rows. Since we have to apply gates from the first stage before we can apply gates from the second stage (to the same column), we can't take advantage of many of the local interactions made available during a single iteration of the swap network: we have to wait for every vertically adjacent qubit from the 2\textsuperscript{nd} and 3\textsuperscript{rd} rows to move to the local interaction zone and have a Givens rotation applied to them before we can take advantage of any of the other local interactions made available. In short: we lose the ability to parallelise, but gain something from not needing to implement the $Z$-strings for every vertical interaction. 

This problem becomes even worse for larger grid sizes, and dramatically worsens the scaling of the algorithm. Table \ref{tab:fft_comparison} provides the depths required to implement the FFT using the above swap network approach. Clearly the depth scales as $O(N)$, compared to the scaling of $O(\sqrt{N})$ for the ancilla-based approach described in the previous section. However, the depth is superior for small grid sizes.

\subsubsection{Modified swap network}\label{sec:tailored_swap_FFT}
It is possible to modify the approach from \cite{kivlichan18} to reduce the complexity even further, using the same approach taken for the implementation of VQE layers described in the main text. In this section we will be using $U_R$ and $U_L$ from Figure \ref{fig:ansatz_order}(b). The basic idea is to repeatedly swap entire columns using parallel FSWAP gates, which eventually allows all vertical interactions to be implemented locally (with respect to the JW encoding).  

To analyse the complexity of this method for implementing the vertical component of the FFT on grids of size $n_x \times n_y$, we can view the swap network as acting on a line (since it swaps entire columns in parallel). Our approach is to apply iterations composed of $n_x/2$ rounds of FSWAP operations, where we alternate between swapping odd-numbered columns with the columns to their right (the operator $U_L$), and swapping even-numbered columns with those to their right (the operator $U_R$). In this way, following the first iteration (i.e.\ after $n_x$ rounds of FSWAP gates), all even-numbered columns have reached (at some point) the left-hand side of the grid, and all odd-numbered columns have reached the right-hand side. This allows us to apply the first round of the FFT on the odd-numbered columns.

After the second iteration, all odd-numbered columns reach the left-hand side, and all even-numbered columns reach the right-hand side. This allows us to apply the first round of the FFT on the even-numbered columns, and the second round of the FFT on the odd-numbered columns (in parallel). We can continue `bouncing' the odd and even columns from left to right in this way until we have been able to apply all $n_y-1$ rounds of the FFT to both sets of columns. This will require $n_y-1$ iterations in total. Since each iteration is composed of $2n_x$ layers of swap operations, the total depth (for the vertical component) will be $2n_x(n_y-1)$ (assuming that the Givens rotations can be implemented in depth 1). Table \ref{tab:fft_comparison} provides the actual circuit depths for implementing the \emph{full} (i.e.\ both horizontal and vertical terms) FFT for a number of different grid sizes.

If we let $T_{\mathcal{F}}(n)$ be the depth of the circuit that implements the 1D FFT on $n$ qubits, then the depth of the circuit required to implement the 2D fermionic Fourier transform using this swap-network approach will be 
\begin{equation*}
     T_{\mathcal{F}}(n_x) + 2n_x \cdot T_{\mathcal{F}}(n_y).
\end{equation*}

\subsection{Summary of approaches to implement the FFT}

In the previous sections we computed the depths of circuits required to implement the FFT using four approaches: a na\"ive implementation, an asymptotically optimal implementation due to Jiang et al.~\cite{jiang2018quantum}, and two swap-network approaches, one of which is due to Kivlichan et al.~\cite{kivlichan18} and the other of which is a novel modification thereof.

The na\"ive approach is immediately seen to be prohibitively costly (in terms of circuit depth) even for smaller grid sizes. The implementation from~\cite{jiang2018quantum}, although asymptotically better, requires relatively high depth circuits for small grid sizes. In addition, the approach requires a number of ancilla qubits, which makes it a less attractive option for implementing the FFT on near-term architectures with few qubits. Finally, we modified a swap-network based approach from~\cite{kivlichan18} to obtain an implementation of the FFT with low circuit depths for smaller grid sizes. The circuit depths for two of the more promising approaches, expressed in terms of the complexity $T_{\mathcal{F}}(n)$ of the 1D fermionic Fourier transform on $n$ qubits, and assuming an arbitrarily connected architecture, are:
\begin{itemize}
    \item Asymptotically efficient implementation (from \cite{jiang2018quantum}): $T_{\mathcal{F}}(n_x) + T_{\mathcal{F}}(n_y) + 2(8n_x + 2n_y - 2)$.
    \item Modified swap-network approach: $T_{\mathcal{F}}(n_x) + 2n_x\cdot T_{\mathcal{F}}(n_y)$\footnote{It should be possible to absorb the horizontal component of the FFT (with cost $T_{\mathcal{F}}(n_x)$) into the FSWAP gates applied during the sorting network, which would reduce the depth of this approach to $2n_x\cdot T_{\mathcal{F}}(n_y)$.}.
\end{itemize}
Hence, if $(2n_x-1)T_{\mathcal{F}}(n_y) < 2(8n_x + 2n_y - 2)$, the swap-network approach will be more efficient. For square lattices of fermions, this condition becomes $T_{\mathcal{F}}(n) < \frac{20n-4}{2n -1 }$. For small $n$, it seems likely that this condition will be satisfied, and therefore the swap-network based approach will be more efficient. Indeed, if $T_{\mathcal{F}}(n) = n-1$ (from the algorithm of~\cite{jiang2018quantum}), this condition is satisfied for $n \leq 11$.

\subsection{Complexity of preparing Slater determinants directly}

Ref.~\cite{jiang2018quantum} describes an approach for preparing Slater determinants on an $n_x \times n_y$ lattice using a sequence of Givens rotations applied to a computational basis state. This work uses a freedom in the representation of Slater determinants, which allows fewer Givens rotations to be applied if the occupation number is known ahead of time.

The circuit derived from this approach runs in depth $n_x n_y - 1$, so is always more efficient than all of the approaches discussed above, apart from the efficient FFT circuit in~\cite{jiang2018quantum}. Compared with the algorithm of~\cite{jiang2018quantum}, the Slater determinant approach will be more efficient for small lattice sizes. Table \ref{tab:fft_comparison} also lists the depths of the circuits required to prepare Slater determinants on various $n \times n$ lattices.
%


\bibliographystyle{mybibstyle}
\bibliography{bibliography}


\end{document}